\documentclass[%
 reprint,
 amsmath,amssymb,
 aps,
prb,
]{revtex4-2}

\usepackage{graphicx}
\usepackage{dcolumn}
\usepackage{bm}
\usepackage{color}
\usepackage[colorlinks,citecolor=blue]{hyperref}

\usepackage[normalem]{ulem}
\definecolor{darkred}{cmyk}{0,1,1,0.2}

\begin{document}

\preprint{APS/123-QED}

\title{The Amaterasu particle: constraining the superheavy dark matter origin of UHECRs}

\author{Prantik Sarmah}
\email{prantik@alumni.iitg.ac.in}
\email{prantiksarmah@ihep.ac.cn}
\affiliation{Indian Institute of Technology Guwahati,
Guwahati, Assam-781039, India}
\affiliation{Institute of High Energy Physics,
Chinese Academy of Sciences, Beijing, 100049, People’s Republic of China}

\author{Nayan Das}
\email{nayan.das@iitg.ac.in}
\affiliation{Indian Institute of Technology Guwahati,
Guwahati, Assam-781039, India}

\author{Debasish Borah}
\email{dborah@iitg.ac.in}
\affiliation{Indian Institute of Technology Guwahati,
Guwahati, Assam-781039, India}

\author{Sovan Chakraborty}
\email{sovan@iitg.ac.in}
\affiliation{Indian Institute of Technology Guwahati,
Guwahati, Assam-781039, India}

\author{Poonam Mehta}
\email{pm@jnu.ac.in}
\affiliation{School of Physical Sciences, Jawaharlal Nehru University, New Delhi-110067, India}

\begin{abstract} 
Amaterasu, the second most energetic ($244$ EeV) cosmic ray particle has been recently detected by the Telescope Array (TA) surface detector. The origin of the TA  Amaterasu event is puzzling, as  its arrival direction points back to a void in the local Universe, lacking conventional astrophysical ultra-high-energy (UHE) cosmic ray sources.  Hence, we explore the possibility if this TA Amaterasu event could have originated from the decay of superheavy dark matter (SHDM) in the Milky Way. Such an origin also opens up multi-messenger detection channels in both  UHE gamma-rays and UHE neutrinos. In this present work, using the TA Amaterasu event and the multi-messenger  limits/sensitivities from various UHE telescopes, we  place stringent constraints on the lifetime and mass of the SHDM. We find that the non-detection of the corresponding gamma-rays at the Pierre Auger Observatory (PAO) and the TA is in severe tension with the SHDM parameter space  required to explain the TA Amaterasu event. Additionally, we  extend the   multi-messenger analysis to the future UHE gamma-ray and UHE neutrino telescopes such as PAO upgrade, GRAND 200k and IceCube-Gen2. We find that the bounds from the future neutrino telescopes  will be able to compete with the present UHECR bounds. However, compared to the existing UHE gamma-ray bounds, the future PAO upgrade and the GRAND 200k gamma-ray detectors will improve the bounds on SHDM lifetime by at least one order of magnitude.

\end{abstract}

\maketitle

\section{Introduction}
The Telescope Array (TA) experiment has recently detected an ultra-high-energy cosmic ray (UHECR) event (named Amaterasu) of energy $2.44 \times 10^{11}$~GeV which is the second highest energy UHECR ever detected~\cite{TelescopeArray:2023sbd}. The arrival direction (R.A.: $255.9 \pm 0.6^{\circ}$, Dec.: $16.1 \pm 0.5^{\circ}$) of this UHECR points back to a void in the local Universe rendering more questions than answers about its origin. As UHECRs above energy $5\times 10^{10}$~GeV propagating more than $\sim 10$~Mpc (mean free path) undergo energy losses due to the Greisen–Zatsepin–Kuzmin (GZK) effect~\cite{PhysRevLett.16.748,1966JETPL...4...78Z,Chakraborty:2023hxp,Gelmini:2005wu,Gelmini:2022evy,Bhattacharjee:1999mup,Stecker:1968uc,Sarmah:2024qww}, 
the likely source of this event is expected to be located within a distance of $\sim 30$~Mpc~\cite{TelescopeArray:2023sbd}. The TA collaboration~\cite{TelescopeArray:2023sbd} analysed a catalogue of gamma-ray sources to identify the probable source and found that the known sources were~\cite{Fermi-LAT:2019yla,Farrar:2008ex} located at distances beyond $30$~Mpc.  
One  possibility could be that the primary UHECR is an iron nucleus at source, which might have undergone significant deflection in the extragalactic magnetic field (EGMF) during propagation to Earth~\cite{Unger:2023hnu,Kuznetsov:2023jfw,Zhang:2024sjp}. Due to this, the actual source direction might be different than the arrival direction.  Magnetic monopole has also been explored in the literature as a possible origin of  the TA Amaterasu event~\cite{Zou:2023ehj,Cho:2023krz,Perri:2023ncd,Frampton:2024shp,Zhang:2024mze,Lang:2024jmc}. Another possible explanation of the origin could be attributed to decaying superheavy dark matter (SHDM) in the Milky Way or any local galaxy along the arrival direction.  In this work, we examine if such a possibility of decaying SHDM in the Milky Way could give rise to the TA Amaterasu  event.

Contrary to the conventional production of UHECRs through the diffusive shock acceleration (bottom-up), dark matter (DM) decay into standard model (SM) particles has also been hypothesised as a top-down UHECR production mechanism~(See review\cite{Bhattacharjee:1999mup}).  In order to produce the observed energy of the TA Amaterasu event via DM decay, energy-momentum conservation states that the DM particles should have mass larger than $\sim 5\times 10^{11}$~GeV and can extend up to the energy scale ($10^{15}-10^{16}$~GeV), close to the typical grand unified theory (GUT) scale~\cite{Berezinsky:1997hy,Birkel:1998nx,Medina-Tanco:1999fld,Dubrovich:2003jg,Kalashev:2007ph,Kalashev:2017ijd,Aloisio:2007bh,Aloisio:2015lva}. Production of such SHDM may be possible via mechanisms such as vacuum or gravitational fluctuations in the early Universe during or after inflation~\cite{Kuzmin:1998kk,Ling:2021zlj,Kramer:2020sbb,Kim:2019udq,Baker:2019ndr,Babichev:2020yeo,Chianese:2020yjo,Supanitsky:2019ayx,Alcantara:2019sco,Babichev:2018mtd,Kolb:2017jvz,Kalashev:2017ijd,Marzola:2016hyt,Kannike:2016jfs,Santillan:2012wq,Arbuzova:2024uwi}. Note that UHECRs can also be produced in principle from self-annihilation of SHDM particles. However, the resultant flux is density squared suppressed~\cite{Murase:2012xs}. Our analysis is based on the Galactic SHDM decay possibility, the extra-galactic component of the flux is found to be negligible in comparison to the  the dominant contribution from Galactic DM halo \cite{Das:2023wtk}. 

SHDM particle ($\chi$) in the Milky Way halo can produce UHECR protons ($p$), ultra-high-energy (UHE) gamma-rays ($\gamma$) and UHE neutrinos ($\nu$) via the decay mode $\chi \to X \bar{X}$. Here, $X$ and $\bar{X}$ are the SM particle-antiparticle pairs~\cite{Gondolo:1991rn,Sarkar:2001se,Barbot:2002gt,Aloisio:2003xj,Bauer:2020jay}. For example, SHDM decaying via  $b$ quark channel  can produce these UHE secondaries: $\chi \to b\bar{b} \to p\bar{p}~\text{or}~ \gamma\gamma ~\text{or}~\nu \bar{\nu}$.  The resulting abundances of gamma-rays and neutrinos generally dominate over that of protons for specific choice of the decay channel. This is in disagreement with the UHECR composition measurements~\cite{PierreAuger:2016use,TelescopeArray:2018bep,TelescopeArray:2020bfv}, that hint towards dominance of nucleon over gamma-rays and neutrinos. Note that abundances of different nuclei from protons to iron in the  UHECR spectrum is still an unresolved issue~\cite{PierreAuger:2016use,TelescopeArray:2018bep,TelescopeArray:2020bfv}. Because of the disagreement with observations, SHDM scenarios for UHECRs are tightly constrained up to the maximum observed CR energy, $\sim 10^{11}$~GeV.  The conventional UHECR spectrum  is heavily suppressed around $10^{11}$~GeV due to the GZK effect and falls with energy as $\sim E^{-5}$~\cite{PierreAuger:2020qqz}. Thus, the region above this energy becomes invaluable for probing scenarios of SHDM.

The UHE gamma-ray and UHE neutrino fluxes associated with the conventional astrophysical UHECR flux are  generally much smaller than that of the UHECR flux~\cite{Chakraborty:2023hxp}. Whereas in SHDM scenarios, we expect them to be larger as mentioned above.  This distinct multi-messenger characteristic can be a useful probe of UHECRs from SHDM and therefore, detection of UHE gamma-rays and UHE neutrinos becomes crucial.  There are mainly two different telescopes, Pierre Auger Observatory (PAO)~\cite{PierreAuger:2021hun,PierreAuger:2016qzd,PierreAuger:2007hjd,PierreAuger:2016use,PierreAuger:2020qqz,PierreAuger:2022jyk,2015NIMPA.798..172P} and TA~\cite{Teshima:1997sc,TelescopeArray:2018bep,TelescopeArray:2018rbt,TelescopeArray:2020bfv,TelescopeArray:2021fpj,TelescopeArray:2023sbd} currently operating and capable of detecting all the three components, UHECRs, UHE gamma-rays, and UHE neutrinos. The Yakutsk Extensive Air Shower Array can detect UHECRs and gamma-rays~\cite{Dedenko:2013rts,Knurenko:2018vdc}, but its sensitivity is lower compared to TA and PAO.  On the other hand, the IceCube experiment is also capable of detecting UHE neutrinos and holds sensitivity similar to that of PAO. The future telescope proposals such as Giant Radio Array for Neutrino Detection 200k (GRAND 200k)~\cite{GRAND:2018iaj} and IceCube-Gen2~\cite{IceCube-Gen2:2020qha} and PAO upgrade~\cite{PierreAuger:2016qzd}  having much larger sensitivities compared to the current ones will provide us with unprecedented opportunities to conduct multi-messenger analysis at these extreme energies.

In this work, we for the first time consider the origin of the recent TA Amaterasu event in light of SHDM decay. We put multi-messenger constraints on the lifetime and mass of the SHDM using 
UHECR, gamma-ray and neutrino telescopes. We also account for the low energy UHECR spectrum detected by the TA. We consider different decay modes for SHDM as well as a wide range ($10^{11}-10^{15}$~GeV) of SHDM mass. 
{\color{black}
The constraints on the lifetime from gamma-ray non-detection limit of TA and PAO are found to be stronger than that of the UHECRs, discrediting  the possibility of the TA Amaterasu event originating from SHDM decay.} Furthermore, we also analyse the potential of the future telescopes mentioned above for probing SHDM lifetime through UHE gamma-ray and UHE neutrino detection.  While future  UHE neutrino detection are found to yield constraints comparable to that from UHECRs, UHE gamma-ray detection will lead to a few order of magnitudes improved constraints than that from UHECRs. Our constraints based on the TA Amaterasu  event are consistent with those existing   in the literature~\cite{Das:2023wtk,Murase:2012xs,Chianese:2021htv,Berat:2022iea,Chianese:2021jke,Ishiwata:2019aet,Arguelles:2022nbl,Esmaili:2012us,Munbodh:2024ast,Muzio:2019leu,PierreAuger:2022jyk,Ando:2015qda,Cohen:2016uyg,Blanco:2018esa,LHAASO:2022yxw,Kachelriess:2018rty,IceCube:2022clp,Song:2023xdk,Song:2024vdc,Chianese:2020yjo,Song:2023xdk,Murase:2012rd,Zhang:2009ut,Maity:2021umk}. These works are based on various observations of UHECRs, UHE gamma-rays, and UHE neutrinos, but not the recent TA Amaterasu event.

The paper is organised as follows. In Sec.~\ref{sec:model}, we describe the model adopted for flux calculation . In Sec.~\ref{sec:multi-messenger}, we present the results of our multi-messenger analysis for both present and future telescopes. Finally we summarize and discuss our main results in Sec.~\ref{sec:conclusion}. The dependence of the results on various DM density profiles is discussed in Appendix~\ref{sec:appendixA} and the constraints from the Galactic center are discussed in  Appendix~\ref{sec:appendixB}.


\section{UHE flux from SHDM decay}
\label{sec:model}
In this section, we discuss the production of UHECRs, UHE gamma-rays and UHE neutrinos from the decay of SHDM and estimate the flux of these daughter particles. 
The SHDM particle, $\chi$ is considered to decay via the $1\rightarrow 2$ process ($\chi \rightarrow X \bar{X}$) into various SM particles-antiparticle pairs. These SM particles and antiparticles can further produce other lighter particle-antiparticle pairs. In particular, we focus on the secondaries convenient for detection and further phenomenology, i.e., the UHECR protons ($\text{p}$), the UHE gamma-rays ($\text{$\gamma$}$) and the UHE neutrinos ($\text{$\nu$}$). 

The differential energy spectra, $d N_{i}/dE$ (where, $i \equiv \text{p, $\gamma$, and $\nu$}$) of these UHE final state particles depend on the SHDM mass ($m_{\chi}$), lifetime ($\tau_{\chi}$) and the inclusive decay rate ($\Gamma$) of $\chi \to X \bar{X} \to i + ...$ ~\cite{Bauer:2020jay}. The flux of an observable particle $i$ produced from DM decay depends on the differential energy spectra given by,
\begin{eqnarray}
   \dfrac{dN_i}{dx} = \dfrac{1}{\Gamma_0}   \dfrac{d\Gamma}{dx} (\chi \to i + \ldots)\ .
\end{eqnarray}
Here, $x = 2 E/ m_{\chi}$ and  $\Gamma_0 = 1/\tau_\chi$ is the inverse lifetime in the rest frame of $\chi$~\cite{Bauer:2020jay}.
The lifetime ($\tau_{\chi}$) and the mass ($m_{\chi}$) of SHDM are considered to be the free parameters of the model and are constrained from the various  observations of the UHE secondaries ($\text{p, $\gamma$, and $\nu$}$).

In particular, the differential energy spectra  ($d N_{i}/dE$) for different decay modes ($\chi \rightarrow X\bar{X} \rightarrow  i + ...$) are computed with the publicly available numerical code \texttt{HDMSpectra}~\cite{Bauer:2020jay}.  \texttt{HDMSpectra}  allows to compute energy spectra for a wide $m_{\chi}$ range, from the electroweak scale up to the GUT scale. The  energy spectra are estimated by the fragmentation function, $D_{a}^{b}(x; \mu_{Q},\mu_{0})$ that gives the probability of an initial state $a$ (e.g., $X \bar{X}$) at energy scale $\mu_{Q}$ evolving to a final state $b$ (e.g., $\rm p + \bar{p}$) at energy scale $\mu_{0}$ with momentum fraction $x=2E/m_{\chi}$~(see \cite{Bauer:2020jay} for details).

\begin{figure*}
    \centering
    \includegraphics[width=0.32\textwidth]{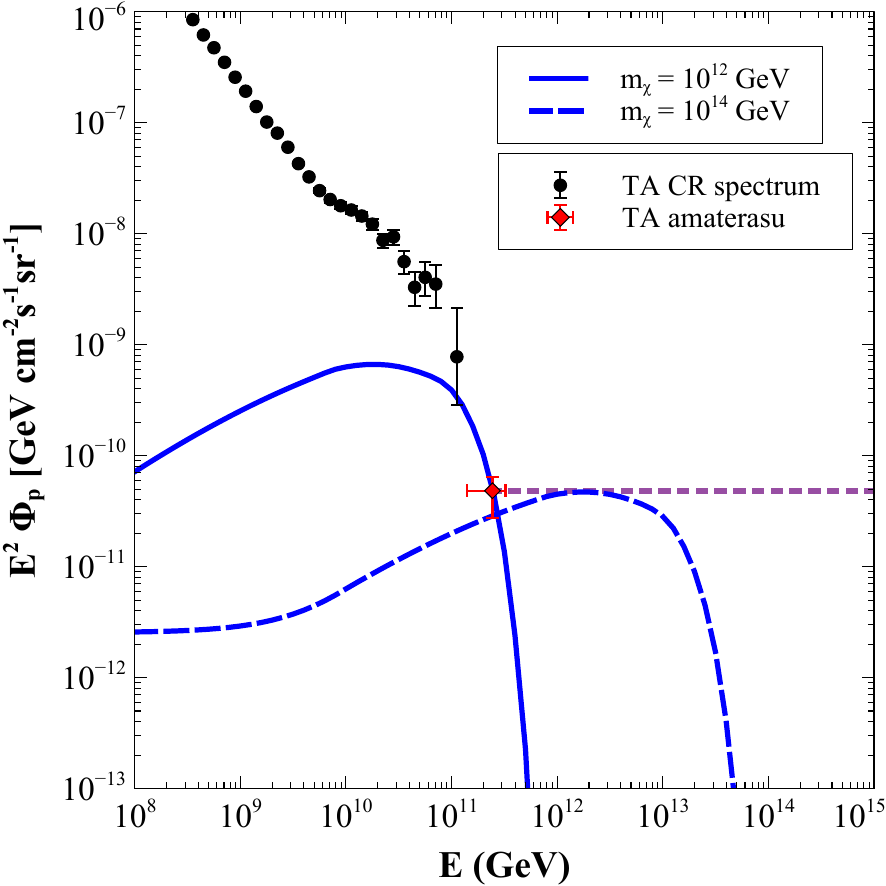}
    \includegraphics[width=0.32\textwidth]{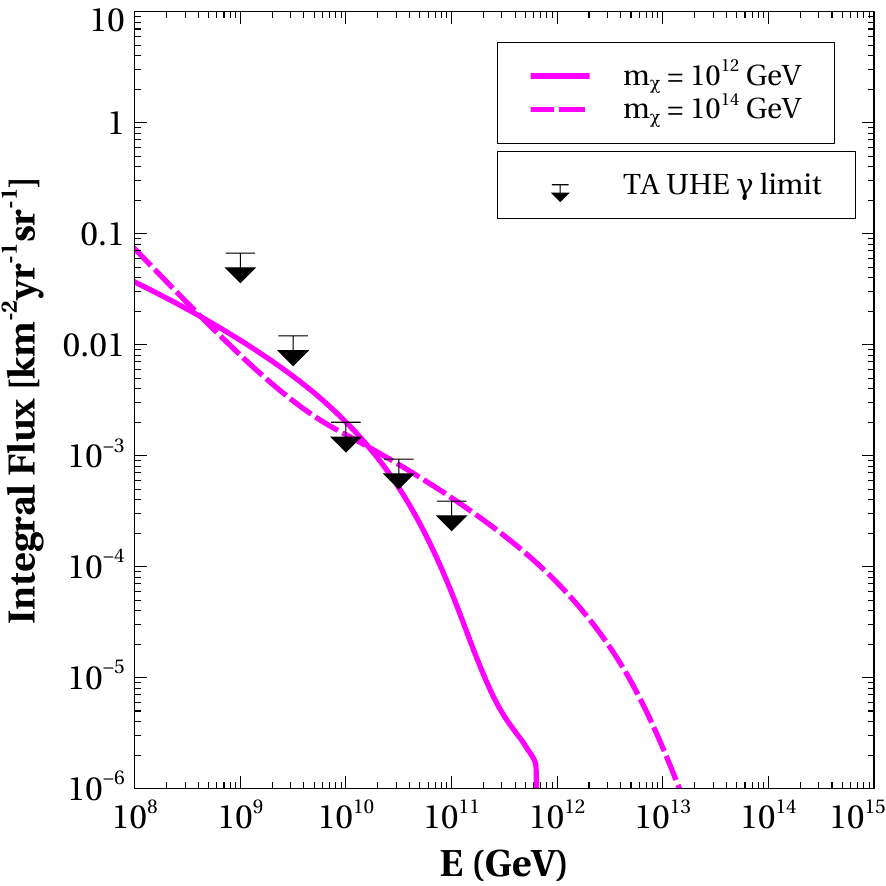}
    \includegraphics[width=0.32\textwidth]{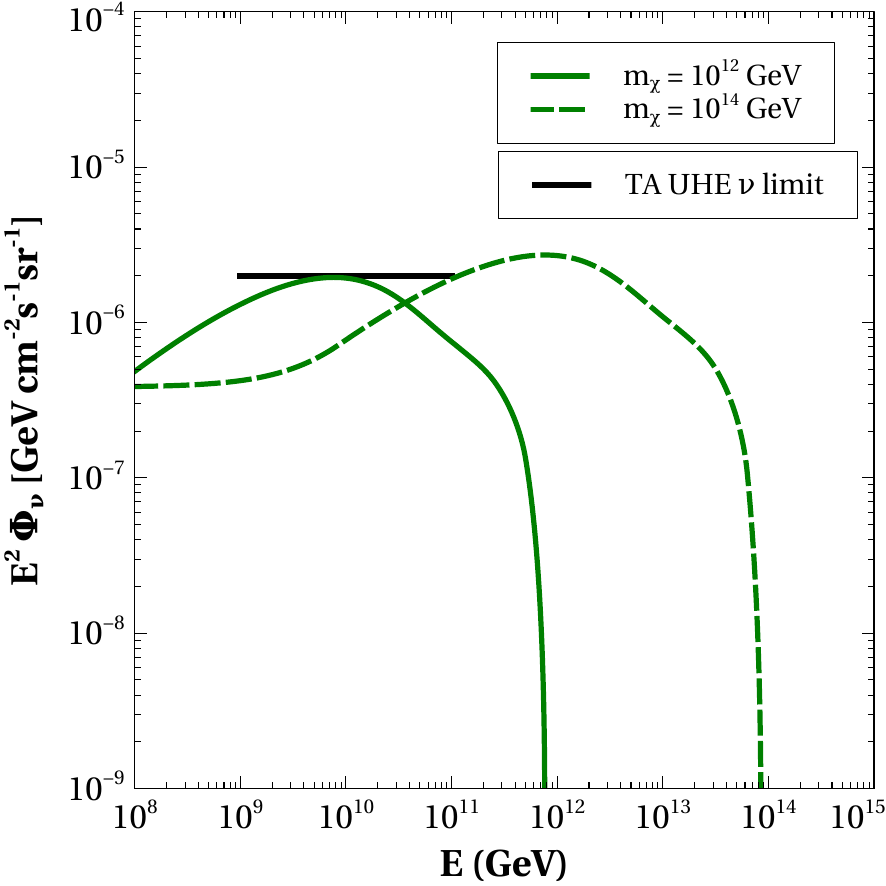}
    \caption{UHE flux  fitted to different observations of TA for $m_{\chi} = 10^{12}$~GeV (solid) and $m_{\chi} = 10^{14}$~GeV (dashed). The decay channel is taken to be $\chi \to b \bar{b}$.
    \textit{Left plot:}  UHECR proton flux (blue curves) from SHDM decay fitted to the  TA Amaterasu event (red data point) and the UHECR spectrum of TA (black data point). The UHECR flux above the energy of the TA Amaterasu event is taken to be constant (purple dotted). The flux fitting resulted in $\tau_{\chi} = 5.5 \times 10^{28}$~s and $7.5 \times 10^{29}$~s for  $m_{\chi} = 10^{12}$~GeV  and $m_{\chi} = 10^{14}$~GeV, respectively. \textit{Middle plot:}   Integral UHE gamma-ray flux (magenta curves) from SHDM decay for the two $m_{\chi}$ values fitted to the upper limit on UHE gamma-ray flux   from TA  (black downward arrow). The resultant values of $\tau_{\chi}$ are $3 \times 10^{30}$~s and $8 \times 10^{29}$~s, respectively. 
    \textit{Right plot:}  UHE neutrino flux (green curves) from SHDM decay for the two $m_{\chi}$ values fitted to the upper limit on UHE neutrino flux from TA (black line). The corresponding values of $\tau_{\chi}$ are $7 \times 10^{25}$~s and $4.8 \times 10^{25}$~s, respectively. The UHE neutrino flux constraint infers significantly small $\tau_{\chi}$ values compared to that of UHE protons and UHE gamma-rays. }
    \label{fig:Fluxes}
\end{figure*}

Now, to estimate the total diffuse secondary particle flux from these SHDM decays one needs to include the contribution from all the SHDM distributed all over the Universe. In principle, the SHDM abundance over the Universe could be crucial. However, it turns out that the extra-galactic contribution is  negligible in comparison to the secondary flux from SHDM decay in the Milky Way~\cite{Das:2023wtk}. Thus, we focus on the diffuse flux of UHE protons, UHE gamma-rays and UHE neutrinos produced from the Galactic abundance of SHDM. The DM density in the Milky Way is expected to be the largest at the Galactic center and decreases  as a function of the Galactocentric distance ($R_{\rm GC}$). 
There are various models and simulations for the density profile of DM in the Milky Way~\cite{Navarro:1996gj,Navarro:2003ew,1989A&A...223...89E,Graham:2005xx,Burkert1995}. In this work, we consider the Navarro-Frenk-White (NFW) profile~\cite{Navarro:1996gj}  given by,
\begin{equation}
    \rho_{\rm NFW}(R_{\rm GC}) = \frac{\rho_{\rm c}}{(R_{\rm GC}/R_{\rm c})(1+R_{\rm GC}/R_{\rm c})^2} \ ,
\end{equation}
where, $\rho_{\rm c}$ is the characteristic density at the characteristic scale $R_{\rm c}=11$~kpc and its value is determined by normalizing $\rho_{\rm NFW}$ with the DM density in the solar neighbourhood, $\rho_{\odot}=0.43~\rm GeV~cm^{-3}$.

Thus,  given mass ($m_{\chi}$) and lifetime ($\tau_{\chi}$) of the SHDM particle, the total flux, $\Phi_{i} (E,\theta)$ of the $i^{th}$ secondary UHE particle ($i= \text{p, $\gamma$, and $\nu$}$) along 
the line of sight can be expressed as,
\begin{equation}
\begin{split}
    \Phi_{i} (E,\theta) 
    &= \frac{1}{4 \pi m_{\chi} \tau_{\chi}} \frac{d N_{ i}(E)}{dE} \int_{0}^{l_{\rm max}(\theta)} \rho_{\chi}(l) dl \\ 
    & = \frac{\rho_{\odot} R_{\rm SGC}}{4 \pi m_{\chi} \tau_{\chi}} \frac{d N_{ i}(E)}{dE} \mathcal{J} (\theta)  \ ,
    \label{eq:SM_flux}
\end{split}
\end{equation}
where, 
\begin{equation}
    \mathcal{J} (\theta) = \frac{1}{\rho_{\odot} R_{\rm SGC}} \int_{0}^{l_{\rm max}(\theta)} \rho_{\chi}(l) dl \ .
    \label{eq:SM_flux_3}
\end{equation}
Here, $R_{\rm SGC}$ is the distance between the Sun and the Galactic center and is taken to be $8.34$~kpc. This allows changing the line of sight coordinate ($l$) to the convenient Galactocentric coordinate ($R_{\rm GC}$).
The coordinate transformation from  $l$ to $R_{\rm GC}$ is given by $R_{\rm GC} = \sqrt{R_{\rm SGC}^2 - 2 l R_{\rm SGC} \cos{\theta} + l^2}$.

The flux over  a solid angle, $\Omega = 2\pi (1-\cos{\theta})$ can be obtained by integrating  out Eq.~\eqref{eq:SM_flux_3} from $\theta=0$ to $\theta=\theta_{\rm max}$. Thus the angle integrated total flux of the $i^{th}$ secondary is, 
\begin{equation}
    \Phi_{i} (E) = \frac{\rho_{\odot} R_{\rm SGC}}{4 \pi m_{\chi} \tau_{\chi}} \frac{d N_{i}(E)}{dE} \mathcal{J}^{\theta_{\rm max}}   \ ,
    \label{eq:SM_flux_4}
\end{equation}
where, 
\begin{equation}
    \mathcal{J}^{\theta_{\rm max}} = \frac{2 \pi}{\Omega} \int_{0}^{\theta_{\rm max}} \sin{\theta} d \theta \mathcal{J} (\theta) \ . \label{eq:Jmax}
\end{equation}


As we are interested in the diffuse flux of UHECRs, UHE neutrinos and UHE gamma-rays, we integrate Eq.~\eqref{eq:Jmax} over the entire solid angle, i.e., $\theta_{\rm max} = \pi$ and  obtain $\mathcal{J}^{\theta_{\rm max}} \approx 2$. The dependence of the fluxes of  UHE secondaries on a specific DM density profile  can be estimated by calculating the $\mathcal{J}^{\theta_{\rm max}}$ factor. Note that, a different choice of DM density profile  other than the NFW profile  leads to a negligible change in the UHE secondary fluxes, see Appendix~\ref{sec:appendixA}.

Now, this diffuse flux of the secondary particles can be obtained for different decay modes of $\chi$, i.e., for different quarks, leptons and bosons for the $X$ and $\bar{X}$ combination (eg., $b\Bar{b}$, $W^{+}W^{-}$, $\mu^{+}\mu^{-}$, $\nu\Bar{\nu}$) in the decay process $\chi \to X \bar{X}$. In particular, the differential energy spectra ${d N_{i}(E)}/{dE}$ is estimated for these different decay modes.  
To demonstrate the flux estimation for the UHECRs, UHE gamma-rays and UHE neutrinos, we consider the example of  $\chi \to b \Bar{b}$ mode using Eq.~\eqref{eq:SM_flux_4} with $100 \%$ branching fraction. The results are depicted in Fig.~\ref{fig:Fluxes} together with the observational data and  constraints of these UHE fluxes from the TA experiment. We consider two specific choices of $m_{\chi}$: $10^{12}$~GeV and $10^{14}$~GeV. The fluxes of all these UHE particles are inversely proportional (see Eq.~\eqref{eq:SM_flux_4} ) to $\tau_{\chi}$. Rather than fixing the parameter $\tau_{\chi}$ arbitrarily, we constrain the value of $\tau_{\chi}$ in a phenomenological manner. $\tau_{\chi}$ is estimated by normalising the SHDM derived flux of UHECRs, UHE gamma-rays and UHE neutrinos with the observed data or experimental upper limits on these fluxes. Since these fluxes as well as the observed flux/limits are all different from each other, they give rise to different values of $\tau_{\chi}$.

The left plot of Fig. \ref{fig:Fluxes} shows the observed UHECR spectrum data by TA in black dots. The red data point represents the flux of the recently observed  TA Amaterasu event of energy  $2.44 \times 10^{11}$~GeV  \cite{TelescopeArray:2023sbd}. The flux for this TA Amaterasu event is obtained by dividing the total number of events by the total exposure of the telescope~\cite{TelescopeArray:2023sbd}. The uncertainties in the flux correspond to the propagated uncertainties. Importantly, for the energies higher than $244$ EeV, we consider  a conservative estimate that the maximum possible CR flux is not larger than the observed TA flux at the $244$ EeV event, shown by the purple dashed line. This is contrary to the harder CR spectra suggestions at the highest energies \cite{PierreAuger:2016use,PierreAuger:2022atd}. For these maximum limits, we show the upper limit of $p+\Bar{p}$ flux for two different DM masses ($m_{\chi}=10^{12}$ GeV and $m_{\chi}=10^{14}$ GeV) considering the decay of DM into $b\Bar{b}$ channel. This establishes the lower bound on DM lifetime, as indicated by Eq. \eqref{eq:SM_flux_4}. For the DM masses of $10^{12}$ GeV and $10^{14}$ GeV, the observational constraints fix the $\tau_{\chi}$ at $5.5\times 10^{28}$ s  and $7.5\times 10^{29}$ s,  respectively. Note that the required energy for the TA Amaterasu event can not be created from SHDM with $m_{\chi} \lesssim 10^{12}$~GeV. This can be understood from the flux corresponding to $m_{\chi}=10^{12}$~GeV in the left plot of Fig.~\ref{fig:Fluxes} as the TA Amaterasu event lies on the high energy tail of  this flux. For $m_{\chi} \lesssim 10^{12}$~GeV, the flux will fall off at energies smaller than the energy of the TA Amaterasu event.  Similarly, there exists an upper limit for the 
SHDM mass. For example, the flux corresponding to $m_{\chi} = 10^{14}$~GeV is falling below the TA flux for our assumption of constant UHECR limit beyond the energy of the TA Amaterasu event.  In particular, we find an upper limit on $m_{\chi}$ for this specific channel to be $m_{\chi} \sim 4 \times 10^{13}$~GeV. However, later in Sec.~\ref{sec:multi-messenger}, we show that this range of $m_{\chi}$  is in significant conflict with the UHE gamma-ray constraints from different telescopes.

In the middle plot of Fig. \ref{fig:Fluxes}, we show the upper limit on the integral UHE gamma-ray flux from the TA experiment~\cite{TelescopeArray:2018rbt,TelescopeArray:2021fpj}.  This upper limit comes from the non-detection of UHE gamma-rays. The solid and dashed magenta lines represent the gamma-ray flux for DM masses of $10^{12}$~GeV and $10^{14}$ GeV, the resultant $\tau_{\chi}$ to maintain the TA gamma-ray upper limits are $3\times 10^{30}$ s and $8\times 10^{29}$ s, respectively. Again, the decay of DM is considered to be into $b\Bar{b}$ final states. These values of $\tau_{\chi}$ are significantly larger than those obtained from UHECR constraints. This is due to the  fact that production of gamma-rays from $b \bar{b}$ annihilation is significantly enhanced compared to that  of protons~\cite{Bauer:2020jay}. Similarly, in the right panel plot of Fig. \ref{fig:Fluxes}, we show the TA's  upper limit (black continuous line) on the UHE neutrino flux from the non-detection of UHE neutrinos in TA \cite{TelescopeArray:2019mzl}. The solid and dashed green lines correspond to the flux of UHE neutrinos for the DM masses of $10^{12}$~GeV and $10^{14}$ GeV, respectively. The lifetimes for these two masses respecting the TA neutrino limits turn out to be   $7\times 10^{25}$ s and $4.8\times 10^{25}$ s, respectively. Due to the poorer sensitivity of the neutrino detectors, the $\tau_{\chi}$ limiting values from UHE neutrinos are at least $3-4$ orders of magnitude smaller than the gamma-ray and UHECR constraints. Thus, the three UHE messengers give rise to different constraints on the SHDM lifetime and mass.

In all the three plots of Fig. \ref{fig:Fluxes}, we do not take into account the contribution from the  conventional astrophysical as well as extragalactic SHDM sources. For UHECRs, the contribution of the extragalactic SHDM component is found to be negligible~\cite{Murase:2012xs,  Das:2023wtk}. Whereas, the contribution of the conventional astrophysical component might be significant in the UHECR spectral energy range below the energy of the TA Amaterasu event, i.e., $\lesssim 10^{11}$~ GeV~\cite{Das:2023wtk}.  However, considering a dominant fraction of UHECRs from the conventional astrophysical origin allows only a little fraction of UHECRs from SHDM. This results in stringent constraints on the production of UHECRs via SHDM decay in the DM mass range  $\lesssim 10^{12}$~GeV and do not yield a conservative description of the constraints on SHDM lifetime.  In order to avoid this, we consider that the UHECRs in this energy range are solely due to SHDM decay. This assumption gives rise to a lower bound on the SHDM lifetime. For UHE gamma-rays, the contribution of the extragalactic component is negligible in the energy range of our interest in comparison to the Galactic SHDM component. This is because UHE gamma-rays from extragalactic sources are attenuated severely due to pair production losses in the cosmic microwave background (CMB) and the extragalactic radio background during propagation to Earth~\cite{Chakraborty:2023hxp,Das:2023wtk}. Note that, the attenuation of Galactic UHE gamma-rays due to the pair production losses is negligible~\cite{Das:2023wtk,Maity:2021umk}.
For UHE neutrinos, the contribution from extragalactic sources can be important as unlike gamma-rays, neutrinos do not suffer any propagation losses. Inclusion of extragalactic sources will result in lowering of the UHE neutrino flux from SHDM decay in the Milky Way which, in turn, will give rise to larger SHDM lifetime.  In order to be consistent with the assumption for UHECRs above, we do not include the extragalactic neutrino flux in our analysis. Due to these assumptions, the constraints obtained  on the SHDM lifetime from different messengers can be comprehended as the most conservative constraints. In the following, we discuss these constraints for present and future telescopes in detail and expand the description to different decay channels.

\section{Multi-messenger constraints on SHDM lifetime}
\label{sec:multi-messenger}

\begin{figure*}
    \centering
    \includegraphics[width=0.32\textwidth]{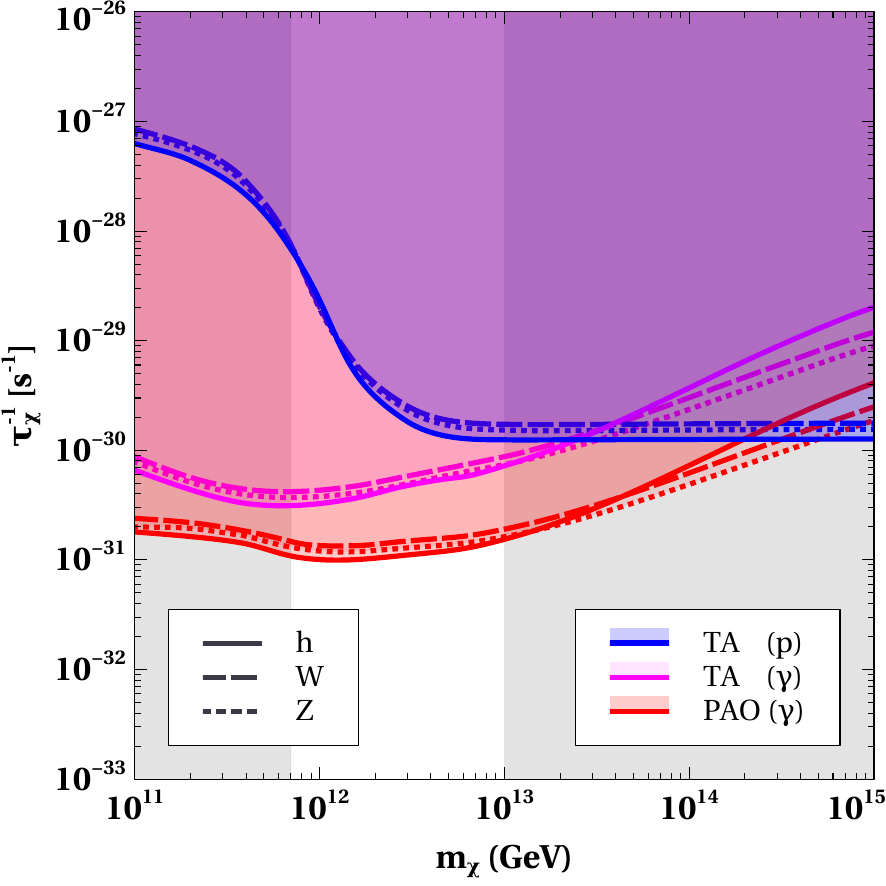}
    \includegraphics[width=0.32\textwidth]{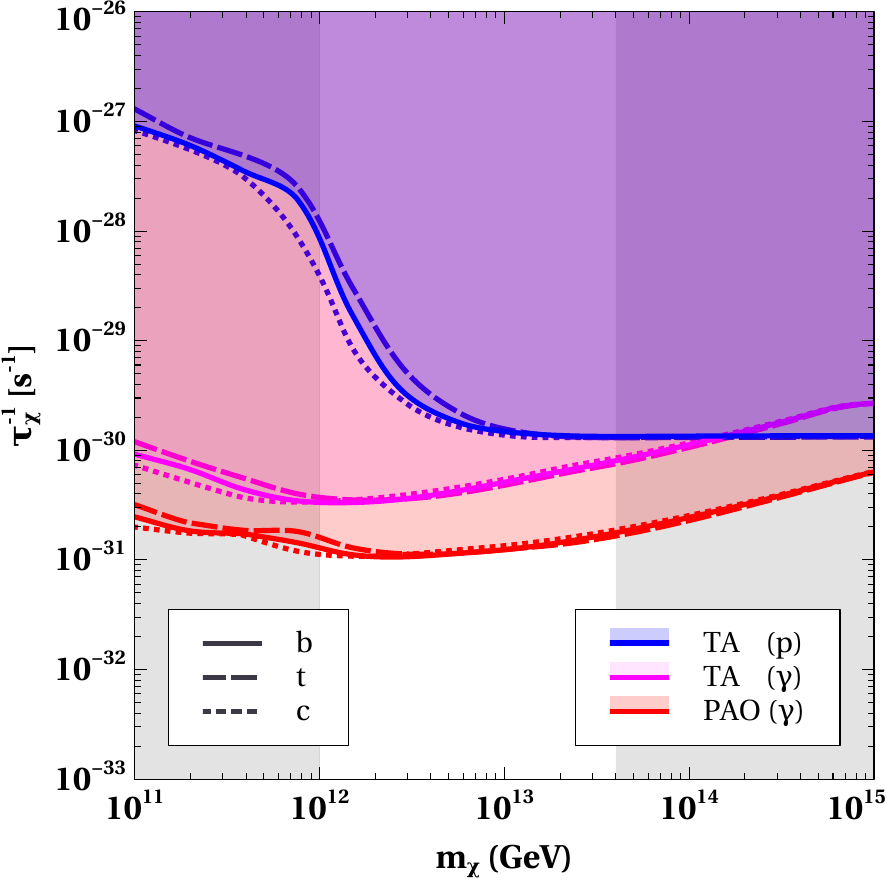}
    \includegraphics[width=0.32\textwidth]{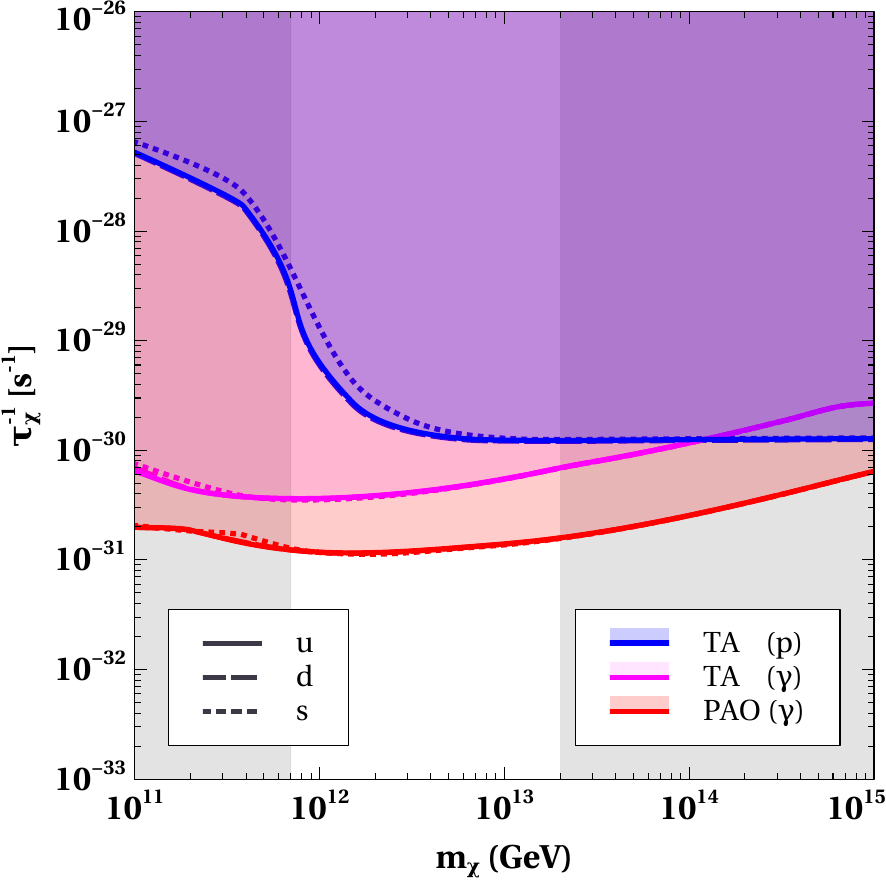}
    \includegraphics[width=0.32\textwidth]{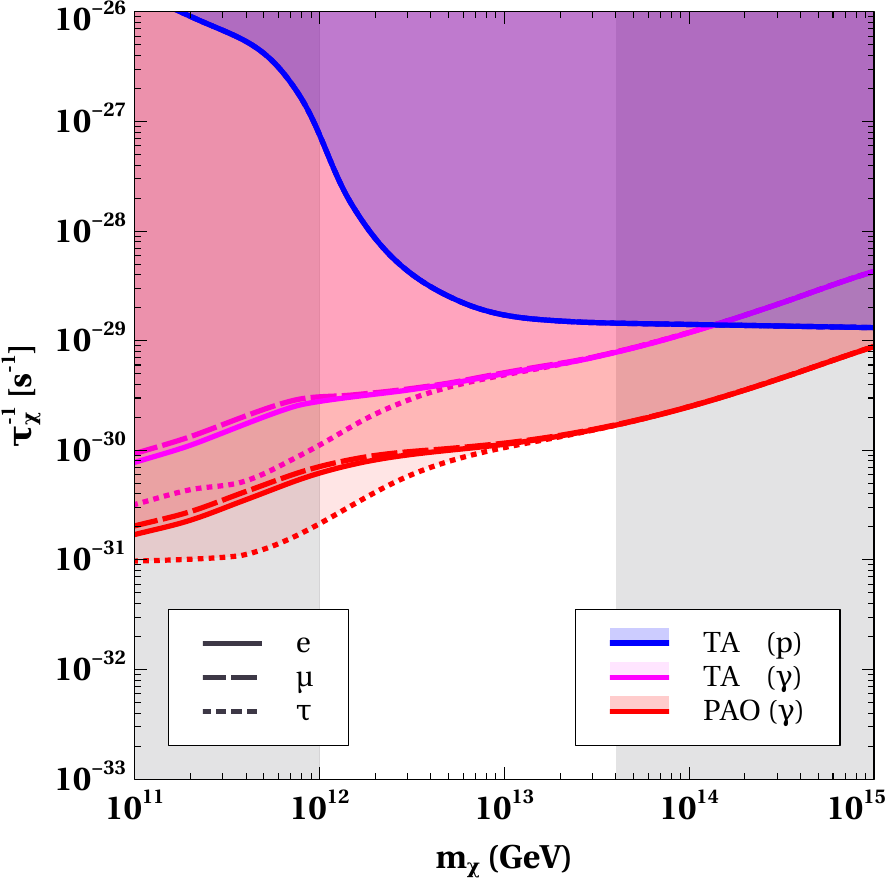}
    \includegraphics[width=0.32\textwidth]{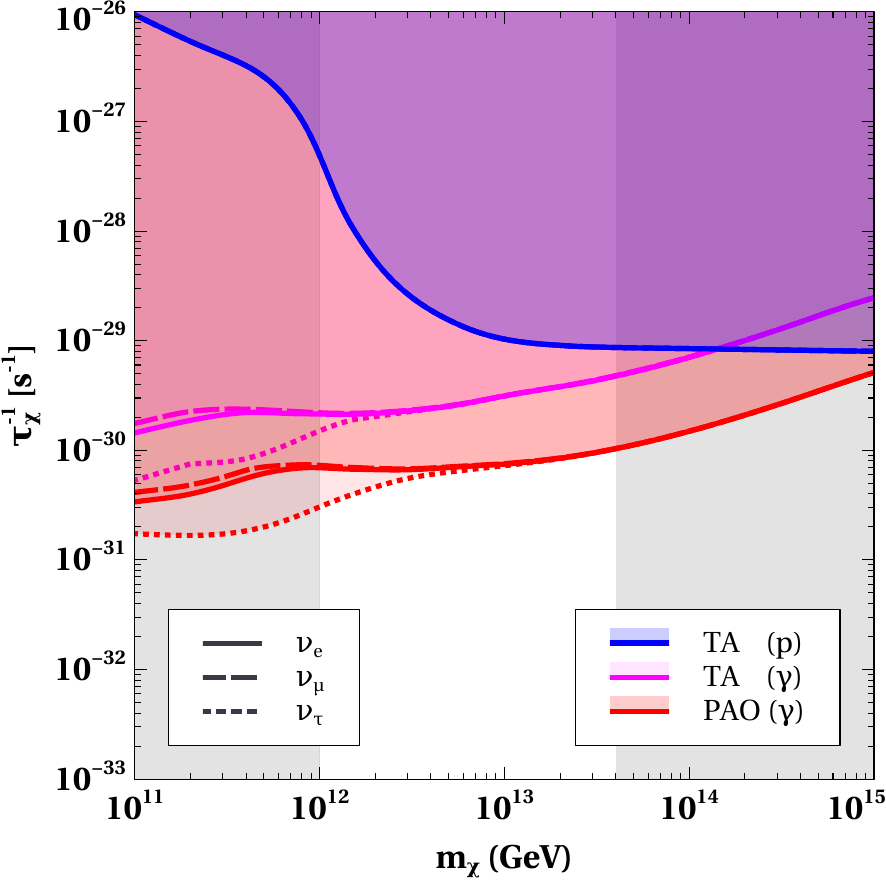}
    \includegraphics[width=0.32\textwidth]{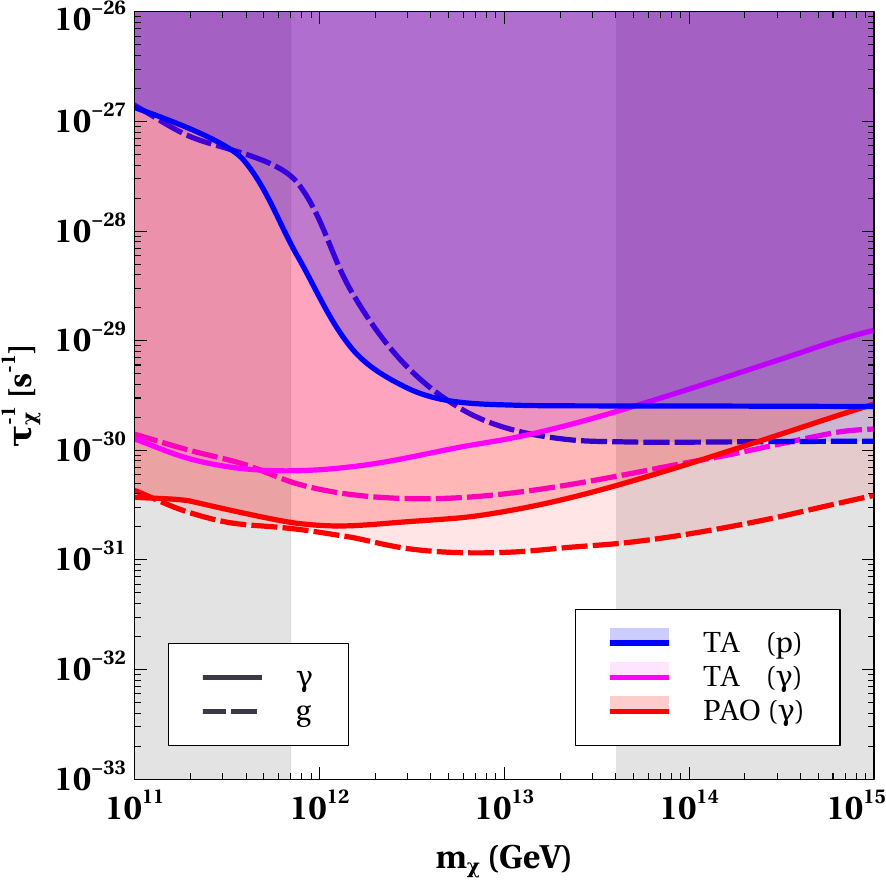}
    \caption{Constraints on SHDM decay rate  for all possible two-body decay channels into SM particles  derived from present UHECR data and UHE gamma-ray non-observation limits. Constraints from UHE neutrino non-observation limits are weak, hence not shown.  The  shaded regions above the curves  in all the plots represent the excluded part of the parameter space. The blue curves show the constraints from the TA Amaterasu event together with the low energy UHECR data. The magenta (TA) and red (PAO) curves depict the constraints from non-observation of UHE gamma-rays. Constraints from different decay modes are distinct. The gamma-ray constraints are in general stronger than that of UHECRs.   The vertically shaded grey regions depict the excluded  range of $m_{\chi}$ for the TA Amaterasu event.}
    \label{fig:current_bound}
\end{figure*}

In this section, we analyse the possibility that the TA Amaterasu event was produced via the decay of SHDM in the Milky Way. Following this, we place constraints on the lifetime/decay rate of SHDM. In order to derive the constraints  in the energy band lower to  the detected energy of the TA Amaterasu event, we consider the UHECR data detected by the TA.  We also carry out a multi-messenger analysis on the SHDM decay rate considering the non-detection of UHE gamma-rays and neutrinos at various telescopes such as TA and PAO \cite{Rautenberg:2021vvt,PierreAuger:2007hjd}. Furthermore, we look into the SHDM lifetime probing capabilities of the various proposed future UHE telescopes. 
Below we present a brief discussion on the  observational capabilities of different past and present telescopes as well as the capabilities of various future telescopes for detection of UHECRs, UHE gamma-rays and UHE neutrinos. \\

{\it UHECR telescopes and measurements:}
UHECRs were discovered by the Akeno Giant Air Shower Array (AGASA)~\cite{2000astro.ph..8102H} and the High Resolution Fly's Eye (HiRes)~\cite{HiRes:2002uqv,HiRes:2007lra} telescopes  about a couple of decades ago. In fact, HiRes had detected the most energetic ($320$~EeV) UHECRs till date~\cite{HIRES:1994ijd}. Recent measurements of the UHECR spectrum above $0.1$ EeV by the PAO~\cite{PierreAuger:2021hun} and TA~\cite{Abbasi:2023swr} experiments agree well with these old measurements except for some specific disagreements regarding mass composition and spectral index at extreme energies. In this work, we focus on the  UHECR spectrum measured by TA and also on the recent TA Amaterasu event. In future, the Giant Radio Array for Neutrino Detection (GRAND)~\cite{GRAND:2018iaj} will complement these existing experiments and may provide us with useful insights into UHECR source properties. \\

\begin{figure*}
    \centering
    \includegraphics[width=0.32\textwidth]{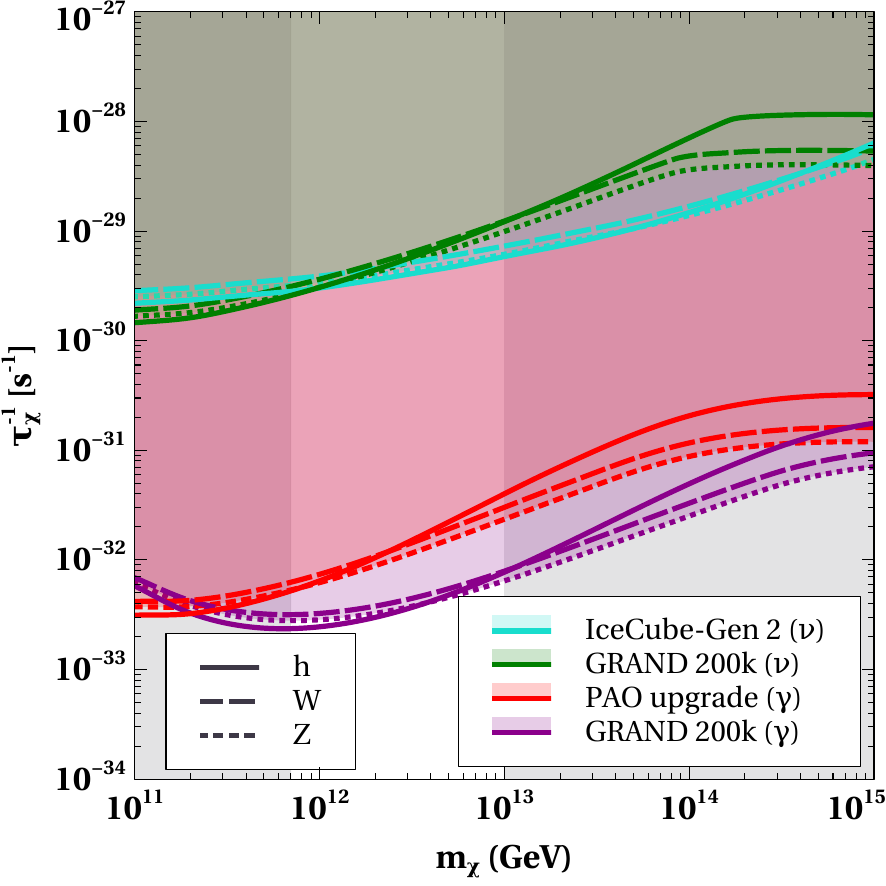}
    \includegraphics[width=0.32\textwidth]{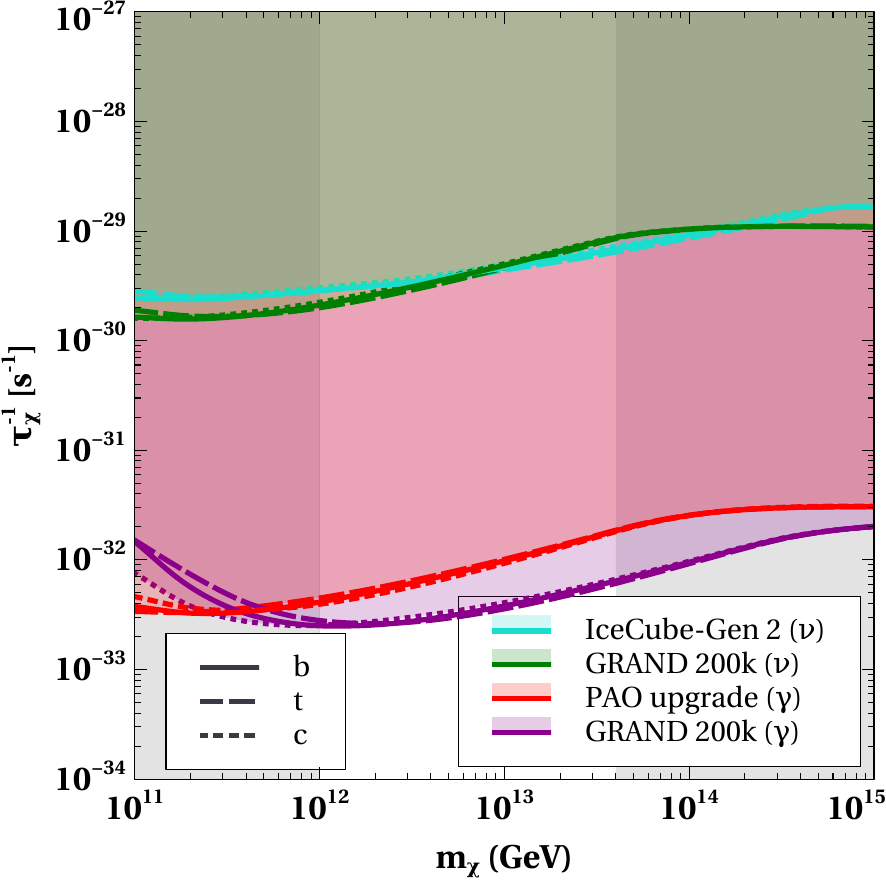}
    \includegraphics[width=0.32\textwidth]{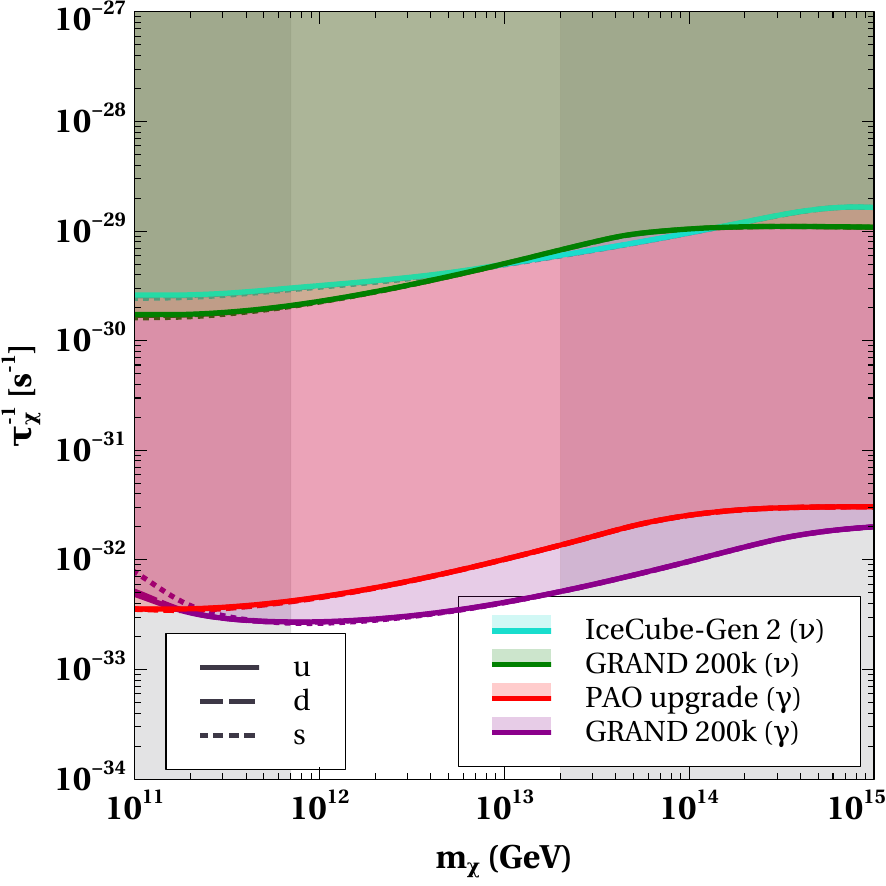}
    \includegraphics[width=0.32\textwidth]{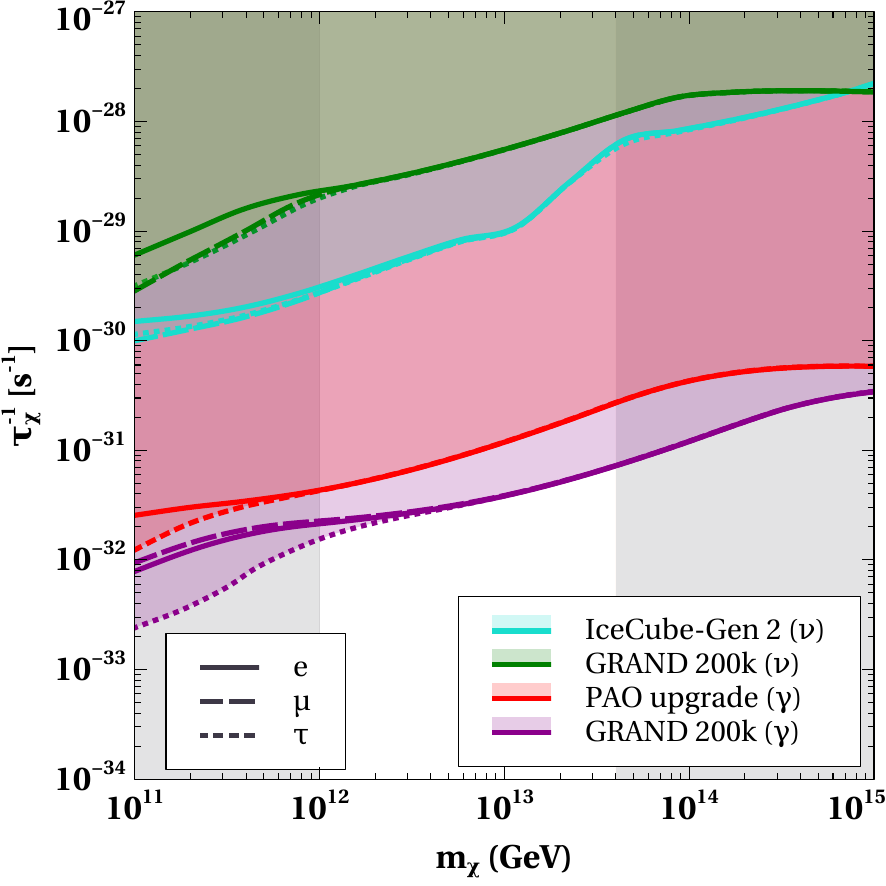}
    \includegraphics[width=0.32\textwidth]{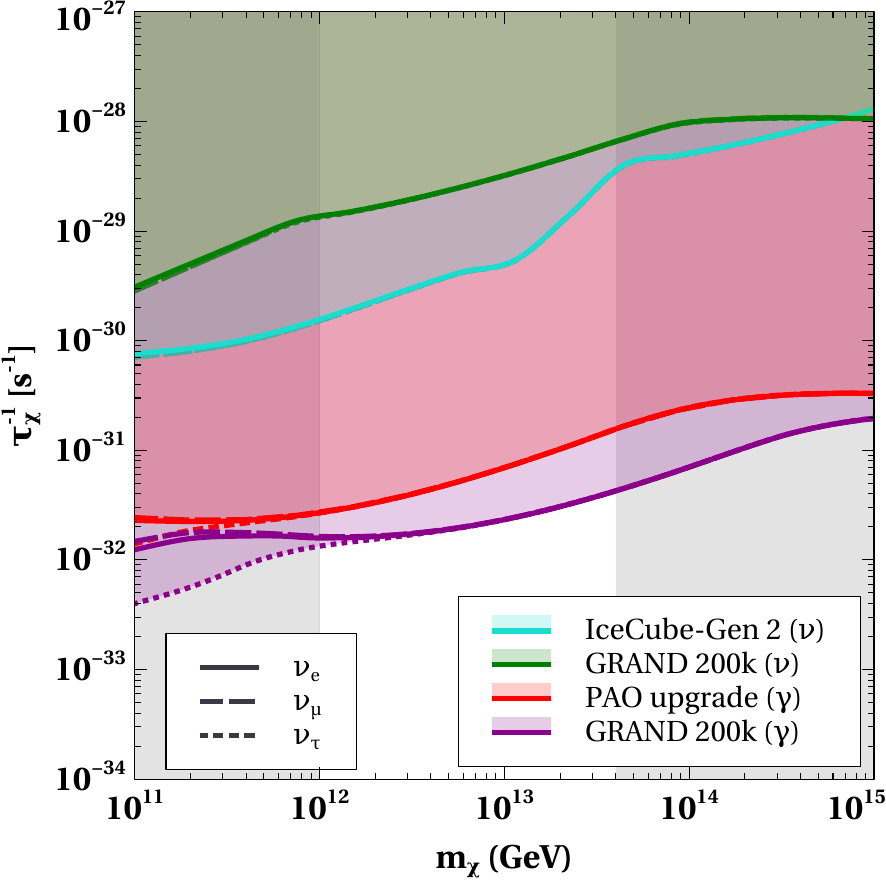}
    \includegraphics[width=0.32\textwidth]{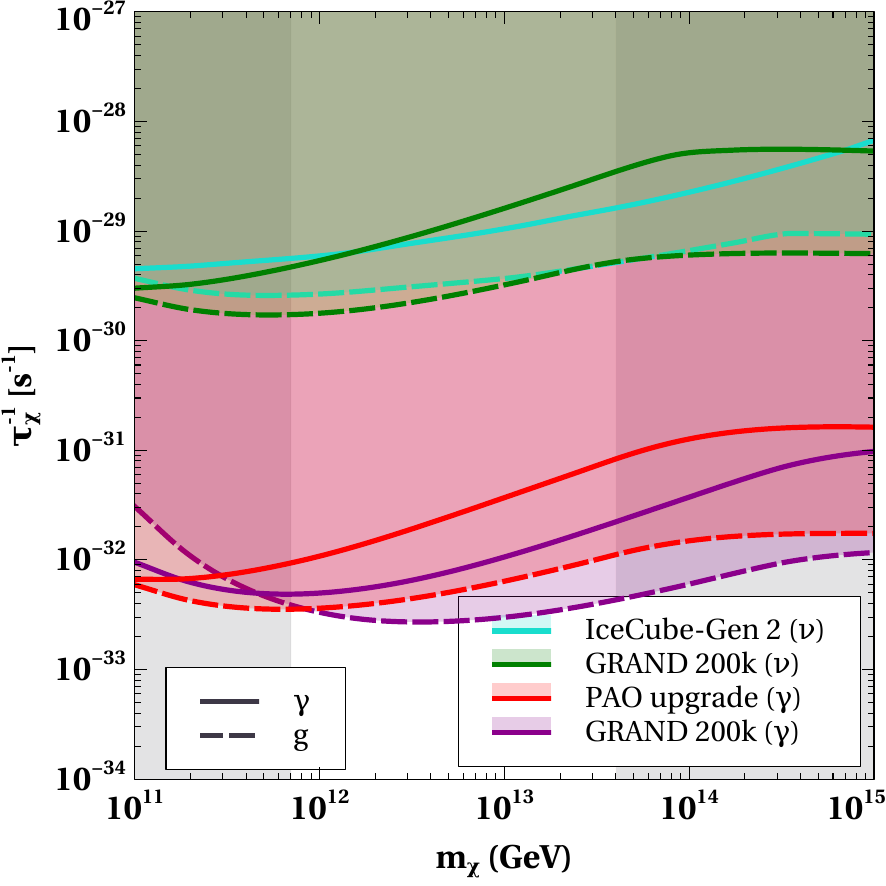}
    \caption{Projected constraints on SHDM decay rate from future observation of UHE gamma-rays and UHE neutrinos for all possible two-body decay channels into SM particles. The  shaded regions above the curves depict the parameter space which can be probed by the future UHE gamma-ray (PAO upgrade: red and GRAND 200k: purple) and UHE neutrino (IceCube-Gen2: cyan and GRAND 200k: green) telescopes.  UHE gamma-ray observations will  put stringent constraints on the SHDM lifetime. The vertically shaded grey regions depict the excluded  range of $m_{\chi}$ for the TA Amaterasu event. }
    \label{fig:future_bound}
\end{figure*}

{\it UHE gamma-ray telescopes:}
There has been no detection of UHE gamma-rays in the PeV-EeV energy band  till date, except for some Galactic sources slightly above PeV energies~\cite{cao2021Natur33C}. Various experiments such as Haverah Park~\cite{Ave:2000nd}, AGASA~\cite{Shinozaki_2002}, Yakutsk~\cite{2007JETPL..85..131G,PhysRevD.82.041101}, AGASA+Yakutsk~\cite{PhysRevD.73.063009,2013ICRC...33..526R}, TA~\cite{TelescopeArray:2018rbt,TelescopeArray:2021fpj}, PAO~\cite{Rautenberg:2021vvt,PierreAuger:2007hjd} have placed different upper limits on  the  UHE gamma-ray flux. Here, we focus on the recent upper limits from TA and PAO.  In future, the GRAND 200k telescopes and  the proposed upgrade of PAO surface detector (SD)~\cite{PierreAuger:2016use} will either detect or  put more stringent constraints on the UHE gamma-ray flux. \\

{\it UHE neutrino telescopes:} Similar to UHE gamma-rays, no UHE neutrinos have yet been detected in the past and presently running detectors such as Antarctic Impulsive
Transient Antenna (ANITA)~\cite{ANITA:2008mzi},  IceCube~\cite{IceCube-Gen2:2020qha}, TA~\cite{TelescopeArray:2019mzl} and PAO~\cite{PierreAuger:2019ens}. Most of the theoretical~\cite{Chakraborty:2023hxp,Kalashev:2002kx,Kusenko:2000fk,Mannheim:1998wp,Madsen:2019hrv,Esmaeili:2022cpz,Li:2007ps,Lee:1996fp,Murase:2014foa,vanVliet:2019nse,Heinze:2019jou,Waxman:1998yy}   and experimental~\cite{IceCube:2018fhm,PierreAuger:2019ens,ANITA:2019wyx,Anker:2019rzo}  upper limits
of the UHE neutrino (astrophysical origin) flux are even below the sensitivities of some of the future telescopes such as Askaryan Radio Array (ARA)~\cite{ARA:2015wxq}, Radio Ice Cherenkov
Experiment Observatory (RICE)~\cite{2012PhRvD..85f2004K}, and Square Kilometer Array (SKA)~\cite{2015aska.confE.144B,9410921}.
However, the upcoming IceCube-Gen2~\cite{IceCube-Gen2:2020qha} and GRAND 200k~\cite{Kotera:2021hbp,GRAND:2018iaj} experiments have significantly larger sensitivities and hold excellent potential of UHE neutrino detection. For the analysis regarding UHE neutrinos, we mainly focus on these two future experiments.

\subsection{Present UHE telescopes}

In this sub-section, we   present the constraints on SHDM lifetime from different observations of UHECRs, UHE gamma-rays and UHE neutrinos and analyse the origin of the TA Amaterasu event.  The constraints are obtained by normalising the flux of these particles with the experimental data or limits from non-observation. For UHECRs, we consider the spectrum measured by TA together with the recent  TA Amaterasu event as depicted in the left plot of Fig.~\ref{fig:Fluxes}. Including the UHECR data from other experiments such as PAO, AGASA and HiRes will yield similar values of DM lifetime as the CR  spectral shapes from these experiments are also very similar to that of TA.  For the UHE gamma-rays, we take the upper limits on the flux from TA~\cite{TelescopeArray:2018rbt,TelescopeArray:2021fpj} and PAO~\cite{Rautenberg:2021vvt,PierreAuger:2007hjd}. Whereas for the UHE neutrinos, we consider the upper limits on the flux from PAO~\cite{PierreAuger:2019ens} and  IceCube~\cite{IceCube-Gen2:2020qha}. We consider a wide range of $m_{\chi}=(10^{11}-10^{15})$~GeV and different two-body decay modes.   For the decay modes, we take  all the quarks (u, d, s, b, t, c), the gauge bosons (W, Z, $\gamma$, gluon), and the Higgs boson (h). The lower limit of $m_{\chi}$ is chosen to be $10^{11}$~GeV as our primary goal is to analyse the origin of the  TA Amaterasu event at $2.44\times10^{11}$~GeV. On the other hand, the upper limit of  $m_{\chi}=10^{15}$~GeV is taken to be close to the GUT scale. For the given $m_{\chi}$, we constrain the lifetime such that the UHE secondary particle observations, limits and sensitivities are satisfied. In particular, we follow convention of exclusion plots and constrain the inverse of the lifetime, i.e., $\tau_{\chi}^{-1}$ rather than the lifetime ($\tau_{\chi}$). The constraints on $\tau_{\chi}^{-1}$  as a function of $m_{\chi}$ for different decay modes are depicted in Fig.~\ref{fig:current_bound}. The shaded regions above the curves depict the excluded part of the parameter space. The vertically shaded gray regions depict the excluded  range of $m_{\chi}$ for the TA Amaterasu event as discussed in Sec.~\ref{sec:model}.

 In Fig.~\ref{fig:current_bound}, the constraints on $\tau_{\chi}^{-1}$ from TA's UHECR data, TA's gamma-ray flux upper limit and PAO's gamma-ray flux upper limit  are shown by the blue, magenta and red curves, respectively.  As the bounds from non-observation of UHE neutrinos are orders of magnitude smaller than the bounds from UHECRs and gamma-rays, the neutrino bounds are not shown in Fig. \ref{fig:current_bound}.
In the top panel, the left plot shows the bounds on SHDM lifetime obtained from  $\chi \to hh,~ W^+W^-, ~ ZZ$, the middle plot shows the bounds from $\chi \to b\bar{b},~ t \bar{t},~ c \bar{c}$, and the right plot shows the bounds from $\chi \to u\bar{u},~ d \bar{d},~ s \bar{s}$. Similarly, in the bottom panel, the left plot displays the bounds from the decay modes $\chi \to e^+e^-,~ \mu^{+} \mu^{-},~ \tau^{+}\tau^{-}$, the middle plot from $\chi \to \nu_{e} \bar{\nu}_{e},~ \nu_{\mu} \bar{\nu}_{\mu},~ \nu_{\tau} \bar{\nu}_{\tau}$, and the right plot from $\chi \to \gamma \gamma, ~ gg$. 
 The excluded regions of the $\tau_{\chi}^{-1}$  vs $m_{\chi}$ parameter  in the plots correspond to flux larger than the experimental data or limits and hence disallowed. 
 
 There are some general features of all these exclusion plots coming for different decay channels. The bounds on $\tau_{\chi}^{-1}$  from UHECR data for $m_{\chi} \lesssim  10^{13}$~GeV mimic the spectral shape  of the UHECR spectrum for all decay modes, whereas above this energy they remain almost constant due to the constant flux assumption above energy of the TA Amaterasu event. If we consider the TA Amaterasu event only, the resulting constraints remain almost constant for all values of $m_{\chi}$ and  are same as the constant values of $\tau_{\chi}^{-1}$ for  $m_{\chi} \gtrsim  10^{13}$~GeV. 

 The different channels also show some distinctions. The constraints from leptons are in general weaker than that of quarks or the bosons by about an order of magnitude. Therefore, the bounds on SHDM lifetime from the TA Amaterasu event can vary depending on the decay mode. However, within a specific family (quarks, leptons, bosons), the bounds do not differ significantly.
 
 Regarding the constraints from UHE  neutrinos,  the present UHE neutrino telescopes such as IceCube, TA and PAO are not sensitive to the flux corresponding to $\tau_{\chi}$ inferred by the TA Amaterasu event and therefore remains consistent  with non-detection of UHE neutrino counterpart to the TA Amaterasu event.  Interestingly, the gamma-ray sensitivities of the TA and PAO are close to the TA Amaterasu constrained SHDM gamma-ray flux. However, no such correlated gamma-ray flux has been detected by the TA and PAO collaboration. This results in much stronger constraints from the gamma-ray upper limits of these telescopes than that of the UHECR data.
 Nevertheless, it is noteworthy that at the extreme values of $m_{\chi}$, close  to $10^{15}$~GeV for the quark channels the UHECR (Fig. \ref{fig:current_bound}, top middle and right plot) constraint is the strongest.

In all decay modes, for values of $m_{\chi}$ above $ \sim 10^{13}$~GeV, the TA UHECR  constraint curves cross the UHE gamma-ray constraint ones, opening a tiny parameter space supporting the SHDM origin of UHECRs. However, this inference is in conflict with the gamma-ray constraints from PAO except for the top left plot 
(bosonic decay modes) in Fig.~\ref{fig:current_bound}. In this case, a small region 
above $m_{\chi} > 10^{14}$~GeV is still allowed for UHECRs of SHDM origin. However, the allowed range of 
$m_{\chi}$ for the TA Amaterasu event is found to be between $\sim (10^{11.8}-10^{13.6})$~GeV. Note the 
slight differences in the allowed mass ranges for different decay modes in the vertically shaded grey bands. The gamma-ray constraints being the strongest in this $m_{\chi}$ range, the SHDM origin of the TA Amaterasu event is found to be in significant conflict with the gamma-ray constraints.  Whereas, outside this range the SHDM decay can not explain both the TA Amaterasu event and the UHECR constant flux limit. Because, in this $m_{\chi}$ range, the constraints from the gamma-ray limits of TA and PAO disfavour the SHDM origin of TA  Amaterasu event.   

While considering the PAO limits one needs to be careful with the field of view (FOV) of the PAO (RA: $0~\text{to}~2\pi$, Dec: $-\pi/2~\text{to}~+\pi/4$). Indeed, the possible coordinate of the recent TA Amaterasu event (RA: $256^{\circ}$, Dec: $16^{\circ}$) is outside the FOV of PAO and can explain the non-detection of any correlated gamma-ray in PAO. In this scenario, the  gamma-ray non-detection constraints from the TA turn out to be more crucial than the PAO constraint. Nevertheless, the values of $\tau_{\chi}$ inferred by the  TA Amaterasu event produce UHE gamma-ray flux larger than the present upper limit from the PAO. Hence, taking into account these multi-messenger constraints, we  argue that the SHDM decay origin of the TA Amaterasu event is in severe tension. 

The constraints extrapolated to higher $m_{\chi}$ values from the TA Amaterasu event   allow a parameter space of  $10^{-30} s^{-1}\lesssim \tau_{\chi}^{-1} \lesssim 10^{-29} ~\rm s^{-1}$ for  $m_{\chi} \gtrsim  10^{14}$~GeV, for a SHDM originated UHECR flux at the extreme energies of the spectrum.  
However, this UHE secondary flux being a diffuse flux, the PAO should also observe gamma-rays at similar energies. For instance, if we consider decay via the $b\bar{b}$ channel, the PAO can be expected to detect a UHE gamma-ray flux of about $(10^{-3}-10^{-4})$ $\rm km^{-2}~yr^{-1}~sr^{-1}$ at $\sim 10^{20}$~GeV. Hence, future detection of UHECRs at or above this energy will be crucial for constraining UHECRs from SHDM decay. We also study the dependence of these constraints on the choice of Galactic DM density profile and find that the results are nearly independent of density profile choices  (see Appendix~\ref{sec:appendixA}). Large DM density at the Galactic center can yield additional constraints on the Amaterasu's SHDM origin. However, these constraints are weak in comparison to those from the diffuse UHECR flux due to certain uncertainties which are discussed in detail in Appendix~\ref{sec:appendixB}.

\subsection{Future UHE telescopes}

The above analysis shows that the observed UHECR data together with the  experimental upper limits on the UHE gamma-ray and UHE neutrino flux put a strong constraint on the SHDM origin for the recent TA Amaterasu  event.  However, these stringent constraints do not completely rule out the possibility of a SHDM component in the UHECR spectrum. It might be possible that the SHDM component of the UHECR flux is smaller than that of the  observed spectrum, eradicating detection in current telescopes and, therefore motivates us to look at the possibilities of the future telescopes. Detection of such a small flux requires more sensitive telescopes. The upcoming telescopes such as GRAND 200k, IceCube-Gen2 and PAO upgrade having larger sensitivities compared to the existing ones provide us with better opportunities to probe the SHDM through detection of  UHE gamma-rays and UHE neutrinos. In the following, we discuss the potential of these future telescopes for constraining  $\tau_{\chi}$ and $m_{\chi}$ for the various two-body decay modes as considered above.

\begin{figure}
    \centering
    \includegraphics[width=0.49\textwidth]{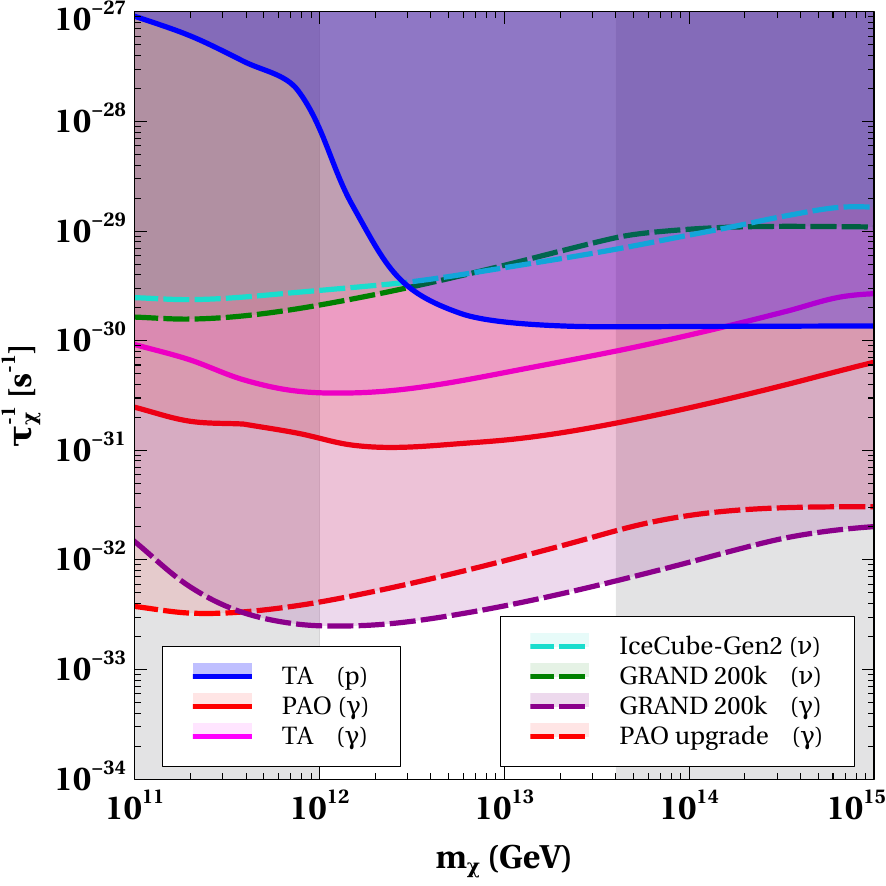}
    \caption{Constraints on DM lifetime/decay rate from both current and future experiments considering the DM decay via $b\Bar{b}$ channel. The blue curve depicts the constraint from TA's UHECR data. The cyan and green dashed curves represent the constraints from UHE neutrino sensitivities of IceCube-Gen2 and GRAND 200k, respectively. The magenta and red solid curves show the present constraints from UHE gamma-ray non-observation in TA and PAO, respectively, whereas, the red and purple dashed curves depict the projected reach of the upcoming telescopes PAO upgrade and GRAND 200k, respectively. While the constraints from current observations (blue, magenta, red solid curves) suggest $\tau_{\chi} \gtrsim 10^{30}$~s,  future gamma-ray observations (red  and purple dashed curves) will be able to probe $\tau_{\chi}$ up to $10^{32}$~s. The vertically shaded grey regions depict the excluded  range of $m_{\chi}$ for the TA Amaterasu event. }
    \label{fig:current_future}
\end{figure}

The constraints on the $\tau_{\chi}$ for various decay modes are obtained by normalising the computed fluxes from SHDM at  the level of  sensitivities (gamma-ray and neutrino) of the said telescopes. Fig. \ref{fig:future_bound} shows the gamma-ray and neutrino bounds on the DM lifetime and mass for the different decay channels. For the gamma-ray constraints, we consider the sensitivities of the proposed PAO upgrade~\cite{PierreAuger:2016use} and the GRAND 200k detector. The neutrino bounds are based on the UHE neutrino flux sensitivities of the IceCube-Gen 2 and the GRAND 200k detectors. The red and purple curves in Fig.~\ref{fig:future_bound} depict the  constraints from gamma-rays in PAO upgrade and GRAND 200k, respectively. The constraints from neutrinos in IceCube-Gen2 and GRAND 200k are shown by the cyan and green curves, respectively.  Overall, the constraints from the gamma-ray sensitivities are about two to three order of magnitudes stringent than that of the ones from neutrinos.  
Similar to bounds from the present experiments, the more stringent limits come from the quarks and gauge bosons decay channels as depicted in Fig. \ref{fig:future_bound} for future telescopes. For gamma-rays, the constraints from GRAND 200k are found to be stronger than that of PAO for all decay modes.
For neutrinos, the constraints from quarks and bosons channels are found to be similar for both the telescopes. However, for the leptonic channels the constraints from IceCube-Gen2 are stronger than that of GRAND 200k. This is because the energy spectra of the leptonic channels give rise to a peak at the energy close to the value of $m_{\chi}$~\cite{Bauer:2020jay}.  
The constraints anticipated from UHE neutrino detection in both the future telescopes will be comparable to that from current UHECR data for $m_{\chi} \gtrsim 2\times 10^{12}$ GeV. 
Therefore, the results from the future experiments will be crucial to put further and stronger multi-messenger constraints on the  parameter space of SHDM. 


\section{Discussion and Conclusion}
\label{sec:conclusion}

In this work, we have analysed the possibility of the recently detected  TA  Amaterasu event, at  $244$ EeV to have originated from the decay of SHDM in the Milky Way. 
The arrival direction of this UHECR being a void in the local Universe makes it hard to  explain its origin from the conventional astrophysical shock acceleration and opens up possibility of a SHDM origin.
The analysis is carried out with a multi-messenger approach considering UHECR data from TA 
and upper limits on UHE gamma-ray and UHE neutrino flux from different experiments such as TA and PAO. For this, we have estimated the UHECR, UHE gamma-ray and UHE neutrino flux produced by SHDM decay with the NFW DM density profile for various SHDM decay channels and a wide range of SHDM mass. The constraints on these fluxes from the experimental data/limits have resulted  in constraints on the SHDM lifetime or decay rates. 
While we have found an  allowed SHDM mass range of  $\sim (10^{11.8}-10^{13.6})$~GeV for the TA  Amaterasu event, the multi-messenger gamma-ray constraints from  TA and PAO  are in severe tension with the SHDM origin for the said TA Amaterasu event. However, these constraints indicate the possibility of  UHECR flux at higher energies originating from heavier SHDM particles, i.e., $m_{\chi} \gtrsim 10^{14}$~GeV.  
In order to test the viability of such a scenario one would need more sensitive future telescopes such as PAO upgrade, IceCube-Gen2 and GRAND 200k.

Therefore, we have also   looked into the potential of these  future telescopes to constrain the lifetime and mass of SHDM. All the results of our work are summarised in Fig.~\ref{fig:current_future} for the decay mode $\chi \to b\bar{b}$. We show the bounds on $\tau_{\chi}^{-1}$ from UHECRs (blue curve), UHE gamma-rays from TA  (magenta curve), PAO (solid red curve), PAO upgrade (red dashed curve), GRAND 200k (purple dashed curve) and UHE neutrinos  from IceCube-Gen2 (cyan dashed curve), GRAND 200k (green dashed curve). The constraints on lifetime from the TA UHECR data are found to be of the order  of $10^{30}$~s for SHDM mass, $m_{\chi} \gtrsim 10^{13}$~GeV,  whereas at lower masses the lifetime can be smaller. The constraints from gamma-ray upper limit of the TA data vary within an order of magnitude and found to be maximum of $\sim 3\times 10^{30}$~s at around $m_{\chi} \sim 10^{12}$~GeV. The TA constraints (UHECR and UHE gamma-ray) alone 
allow for a small parameter space above $10^{14}$ GeV for the SHDM. However, the gamma-ray upper limit of PAO is about a factor of $3$-$4$ larger and creates tension for the SHDM origin of the TA Amaterasu event. 
The future telescopes will be able to put much stronger constraints ($\gtrsim 10^{32}$~s) on the SHDM lifetime from  gamma-ray observations as depicted by the  purple dashed (GRAND 200k) and red dashed (PAO SD upgrade) curves. However, future UHE neutrino detectors such as IceCube-Gen2 (cyan dashed) and GRAND 200k (green dashed) will place constraints similar to the present UHECR derived limits (blue solid). These results vary substantially between different decay modes, we have done a detailed analysis for the different  decay channels and  
found that the lepton channels to have the weakest constraints (see, Fig.~\ref{fig:current_bound} and Fig.~\ref{fig:future_bound}).

To conclude, we have provided an updated estimate of multi-messenger constraints on the lifetime of SHDM by considering  the recent TA Amaterasu event  in conjunction with the TA UHECR data and non-observation of UHE gamma-rays and UHE neutrinos. While doing so, we have ensured consistency with the existing constraints in the literature~\cite{Das:2023wtk,Ishiwata:2019aet,Chianese:2021jke,Chianese:2021htv,Song:2023xdk,Murase:2012xs,Das:2024bed,Maity:2021umk}. 
Note that the constraints for heavier DM mass, i.e.,  $m_{\chi} > 10^{12}$~GeV are not very robust due to the  lack of  data (UHECRs, UHE gamma-rays and UHE neutrinos) at energies above $\sim 10^{11}$~GeV. For harder UHECR flux anticipated from SHDM~\cite{Das:2023wtk} or conventional diffusive acceleration~\cite{PierreAuger:2016use,PierreAuger:2022atd}, the  resulting constraints on the lifetime for $m_{\chi} \gtrsim 10^{12}$~GeV  will be smaller than the ones obtained in this work.   Nevertheless, such  shorter  lifetimes for the SHDM are not supported by UHE gamma-ray observations and do not impact the conclusion of this work.

\begin{acknowledgments}
PS  thanks Toshihiro Fujii for providing crucial information on the TA observation and Pritam Das, Ranjan Laha for useful comments. 
We thank Subir Sarkar for valuable  comments and for pointing out  important references related to the work. 
This research (SC and PM) was supported in part by the International Centre for Theoretical Sciences (ICTS) for participating in the program - Understanding the Universe Through Neutrinos (code: ICTS/Neus2024/04). 
The work of ND is supported by the Ministry of Education, Government of India via the Prime Minister's Research Fellowship (PMRF) December 2021 scheme. 
The work of DB is supported by the Science and Engineering Research Board (SERB), Government of India grants MTR/2022/000575 and CRG/2022/000603. DB also acknowledges the support from the Simons Foundation
(Award Number:1023171-RC) to visit the International Institute of Physics, Natal, Brazil in May 2024 when part of this work was completed. SC and PS have received funding from SERB projects CRG/2021/002961 and MTR/2021/000540.  PM acknowledges the use of HPC cluster at SPS, JNU funded by DST-FIST.
\end{acknowledgments}

\appendix
\nocite{*}
\label{sec:appendix}
\counterwithin{figure}{section}

\section{Dependence on DM density profile on diffuse flux}
\label{sec:appendixA}

\begin{figure*}
    \includegraphics[width=0.4\textwidth]{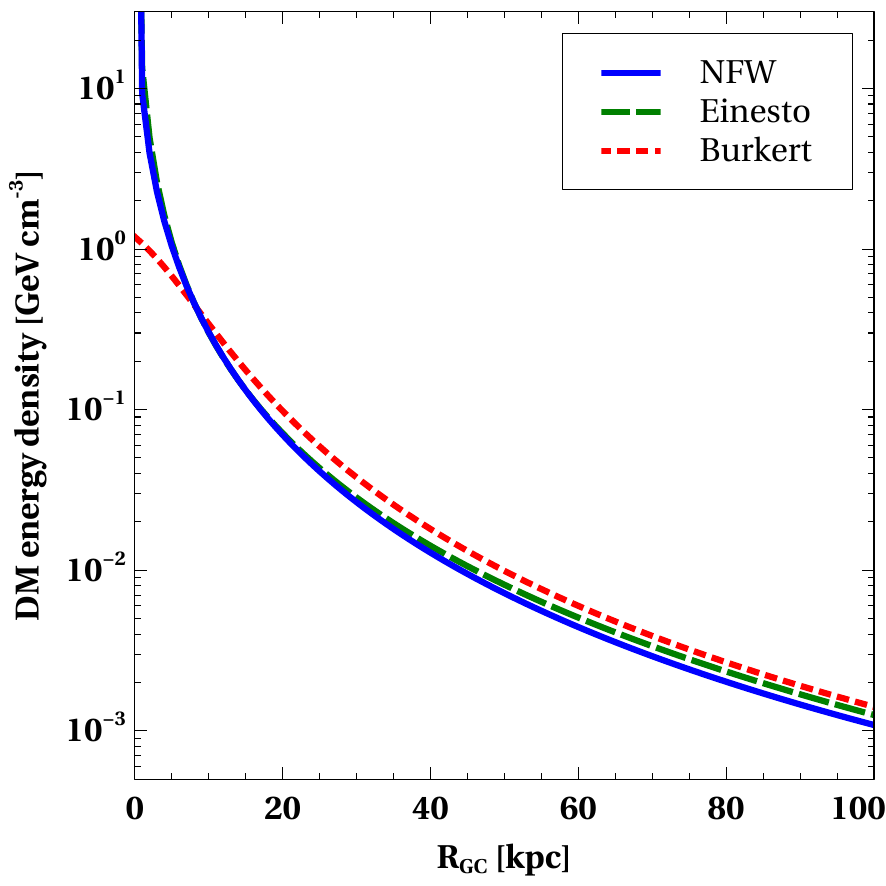}
    \includegraphics[width=0.4\textwidth]{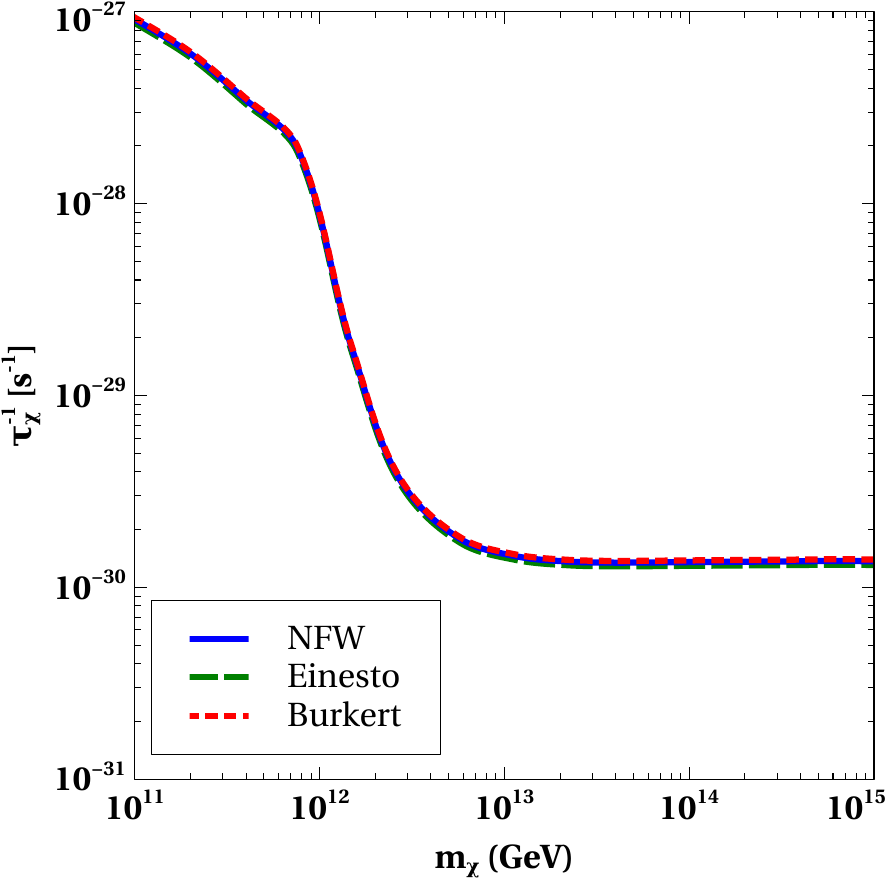}
    \caption{{\it Left plot:} DM energy density vs Galactocentic distance for three different DM profiles. {\it Right plot:} Constraints on DM lifetime from diffuse UHECR flux vs DM mass for different DM profiles considering $b\Bar{b}$ channel. The constraints are nearly independent of the DM profiles. The region above the lines are excluded.}
    \label{fig:DM_profile}
\end{figure*}
We have performed the analysis on SHDM decay considering the NFW DM density profile in the Milky Way throughout the paper. For completeness of the analysis, it is crucial to understand the variation of the results for different DM profiles. Therefore, in this appendix, we present an analysis on the dependence of our results on different DM density profiles. Other than the NFW profile, here we take two other DM profiles - Einesto~\cite{1989A&A...223...89E} and Burkert~\cite{Burkert1995}. The equations for DM energy density for these two profiles are 

\begin{equation}
    \rho_{\rm Ein} (R_{\rm GC})=\rho_{c} \exp \left[- 12. 5\left(\left(\frac{R_{\rm GC}}{R_C}\right)^{0.16} -1\right)\right],
\end{equation}

\begin{equation}
    \rho_{\rm Bur} (R_{\rm GC}) = \frac{\rho_{c} R^3_{c}}{(R_{\rm GC} + R_{c})(R^2_{\rm GC} + R^2_{c})},
\end{equation}
respectively. 

To illustrate the effect of different DM profiles, we compute the constraints on SHDM lifetime from TA's UHECR observation for the DM profiles discussed above. The results are shown in Fig.~\ref{fig:DM_profile}. In the top panel of \ref{fig:DM_profile}, we show 
the DM energy density in the Milky Way as a function of Galactocentric distance for the different DM profiles.  These  DM profiles give different $\mathcal{J}^{\theta_{\rm max}}$ values in equation \eqref{eq:SM_flux_4}. While for NFW profile, $\mathcal{J}^{\theta_{\rm max}}$ takes value of $1.99$, for Einesto and Burkert profiles, its value is $2.08$ and $1.95$ respectively. This variation of $\mathcal{J}^{\theta_{\rm max}}$ gives a maximum of around $7\%$ changes in SHDM lifetime depending on the DM profile as shown in the bottom plot of figure \ref{fig:DM_profile}. Hence, the dependence of the constraints on the DM profile is negligible. 

\section{Galactic Center contribution}
\label{sec:appendixB}

\begin{figure*}
    \centering
    \includegraphics[width=0.4\linewidth]{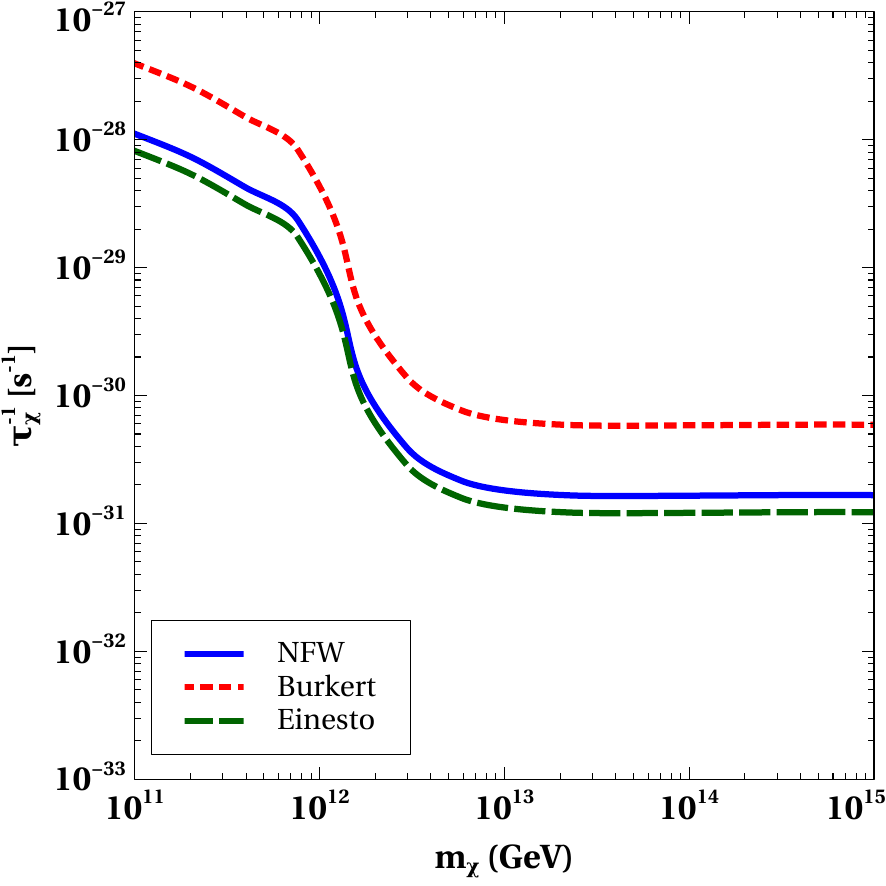}
    \includegraphics[width=0.4\linewidth]{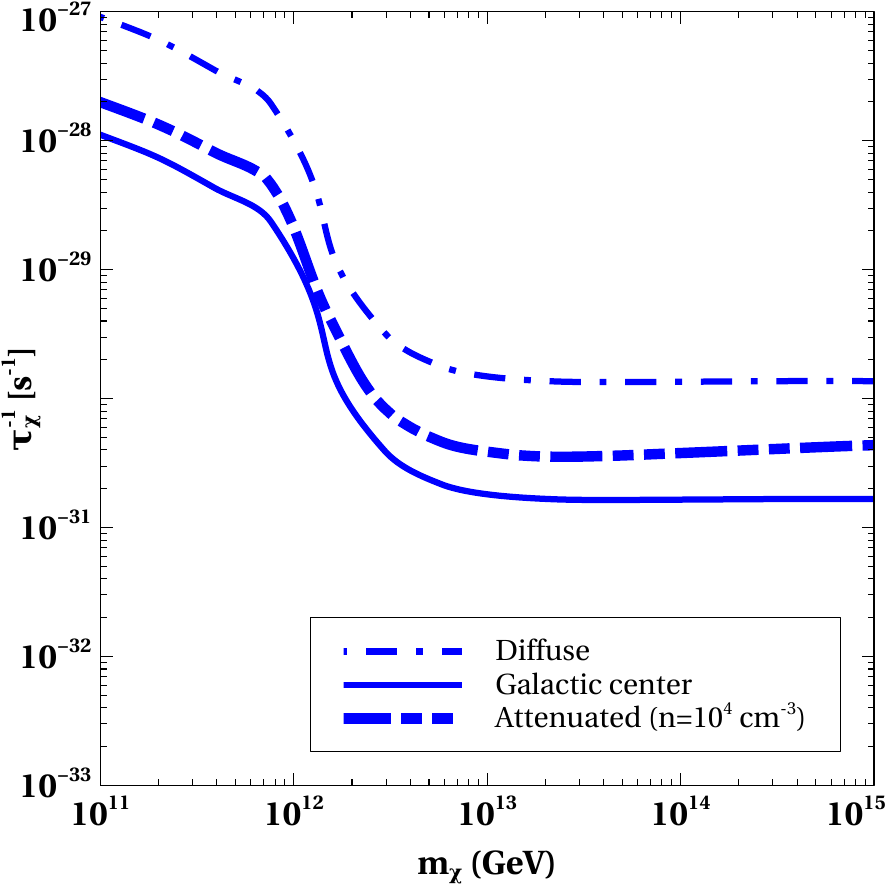}
    \caption{ {\it Left plot :} constraints on SHDM lifetime from the direction of Galactic center for three different DM profiles : NFW (solid blue), Burkert (dotted red) and Einesto (dashed green). {\it Right plot :} Variations on constraints due to attenuation of UHECR from the GC shown by solid blue (without attenuation) and thick dot-dot-dashed blue (attenuated with $n_{\rm CMZ} \sim 10^{4}$ cm$^{-3}$)  lines. For comparison, diffuse flux bound is plotted with dot-dashed blue line. The DM profile  considered here  is the  NFW profile.}
    \label{fig:GC-DM}
\end{figure*}

\begin{figure*}
    \centering
    \includegraphics[width=0.4\linewidth]{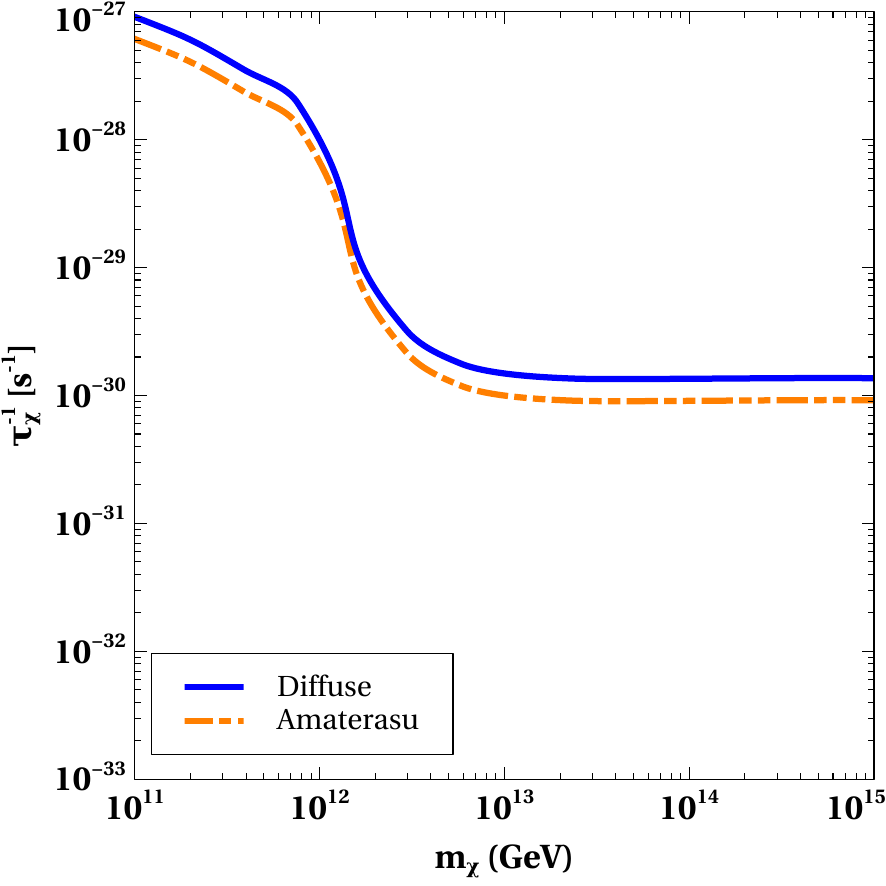}
    \caption{Constraints on SHDM lifetime from the direction of Amaterasu (orange dot-dot dashed) compared with that from the diffuse flux (blue solid) for NFW DM profile. }
    \label{fig:Diffuse_DM1}
\end{figure*}

In the above calculations, we focus on the diffuse part of the secondary fluxes originating from  SHDM decay in the Milky way.  
However, significant anisotropy~\cite{PhysRevD.74.023516} in these secondary fluxes is expected due to our location away from the Galactic center (GC). The anisotropy is expected due to the large DM density in the Galactic center. Indeed, this DM density at the GC can vary for different DM profiles.  To demonstrate this, we again consider the Einesto~\cite{1989A&A...223...89E} and the Burkert~\cite{Burkert1995} models, together with the NFW profile. 
To estimate this contribution  
from the Galactic center direction, we consider a solid angle defined by the Galactic latitude ($-5^{\circ},5^{\circ}$) and longitude ($0^{\circ},5^{\circ}$). The resulting constraints on the SHDM lifetime for the three DM profiles are shown in the left panel of Fig.~\ref{fig:GC-DM}. The Einesto's (dashed green) and Burkert's (dotted red) DM profiles yield the strongest and weakest constraints, respectively. The constraint from the NFW profile (solid blue) remains slightly weaker than those from the Einesto profile. Clearly, in comparison to the DM model dependence of the diffuse flux (Fig.~\ref{fig:DM_profile}) the GC flux shows a large uncertainty and hence not the preferred choice for phenomenology. 

{\color{black}
In addition to the DM model uncertainties, the flux from the GC region might be affected by the associated attenuation of UHECRs due to the large density of molecular cloud at the Milky Way center. The flux from the Galactic center would suffer attenuation due to the interaction (e.g., inelastic $pp$ collision) with the dense molecular gas in the ``Central Molecular Zone (CMZ)". Depending on the density, the CMZ extending up to the radius of about $R = 300$ pc~\cite{Mills} can cause substantial attenuation to this secondary fluxes. In particular, the CMZ density ($n_{\rm CMZ}$) can vary in several orders, the difference arises in different length scales around the Galactic center. Typically, the $n_{\rm CMZ}$ is quoted in the range $\sim (10^{2}-10^{3})~\rm cm^{-3}$. However, there are regions of dense molecular gas clouds within the CMZ with density exceeding $10^{3}~ \rm cm^{-3}$~\cite{Mills,2018ApJS..236...40T}. Indeed, in the parsec scale the density of molecular gas clouds in the CMZ can reach up to  ($~\sim 10^{4}-10^{7}~\rm cm^{-3}$) \cite{2023ASPC..534...83H}. To mimic the absorption effect of the high density gas clouds, we extend the upper limit of CMZ density to  $10^{4}~ \rm cm^{-3}$. For these high densities the absorption of UHECRs from GC is found to be non negligible and may lead to larger attenuation than that of the typical CMZ density. Thus, along with the DM model, this uncertainty of the scale and density of the $n_{\rm CMZ}$ adds to the ambiguity of the SHDM flux from GC.}  

{\color{black}
This has been shown in the plot on the right of Fig. \ref{fig:GC-DM} shows the limit on $\tau^{-1}_{\chi}$ from the Galactic center region for NFW profile with decay mode $\chi \to b\bar{b}$. The bound without the CMZ attenuation is shown by the continuous solid blue curve. The bound with attenuation corresponding to $n_{\rm CMZ}= 10^{4} ~\rm cm^{-3}$  is shown by the thick  dot-dot dashed blue curves. We also show the diffuse flux bound by the dashed-dot blue curve for comparison. 
This shows that the attenuation of UHECRs from the GC can be significant for the large CMZ density $ n_{\rm CMZ} \sim 10^{4} ~ \rm cm^{-3}$.  The attenuation is found to be negligible for  $n_{\rm CMZ} \leq 10^{3} ~ \rm cm^{-3}$. The attenuation can also be impacted by the composition of the CMZ. For our estimation, we choose one nucleon per molecule to be conservative. Note, that the UHECR protons can also interact with low energy photons (photo-pion production), but the attenuation due to this process is found to be negligible compared to the $pp$ one. 

The uncertainty in the UHECR flux from the GC is dominated by the uncertainty due to the DM  profile, but may also have contribution from the CMZ. 
Therefore, the GC flux is not the best candidate for such phenomenological studies. 
Hence, we focus on diffuse contribution of the SHDM to obtain the bounds on the decay rate. Keeping these model uncertainties aside, the non-discovery of these fluxes in PAO should still be important.  
}

Finally, we also compute the constraints on SHDM lifetime from the direction of the Amaterasu by varying the Galactic coordinates in the uncertainty range of the Amaterasu's arrival direction for the NFW profile. Note that the constraints are almost independent of the choice of DM profile as the Amaterasu's arrival direction is about $45^{\circ}$ away from the Galactic center direction. The resulting constraints are compared with those from diffuse flux and the Galactic center direction (NFW) in Fig.~\ref{fig:Diffuse_DM1}. The constraints from the Amaterasu's direction (orange dot-dot dashed) are slightly stronger (about a factor of $1.5$) than those from the diffuse flux. Therefore, our constraints derived from the diffuse UHECR flux are the conservative ones.




\bibliography{apssamp}{}
\bibliographystyle{apsrev4-1}

\end{document}